\documentclass[fleqn,usenatbib]{mnras}
\usepackage{newtxtext,newtxmath}

\usepackage[T1]{fontenc}
\usepackage{ae,aecompl}

\usepackage{graphicx}	
\usepackage{amsmath}	
\usepackage{amssymb}	
\usepackage{mathtools}  
\usepackage[dvipsnames]{xcolor}

\newcommand\subI{_{_\textsc{\tiny i}}}
\newcommand\subQ{_{_\textsc{\tiny q}}}
\newcommand\subU{_{_\textsc{\tiny u}}}
\newcommand\subV{_{_\textsc{\tiny v}}}
\DeclarePairedDelimiter\abs{\lvert}{\rvert}   
\renewcommand{\d}[1]{\ensuremath{\operatorname{d}\!{#1}}}

\title[PRT, RM and large-scale magnetic fields]{Polarised radiative transfer, rotation measure fluctuations and large-scale magnetic fields}

\author[On et al.]{Alvina Y. L. On$^{1,3}$\thanks{E-mail: alvina.on.09@ucl.ac.uk (AYLO), y.chan.12@ucl.ac.uk (JYHC), kinwah.wu@ucl.ac.uk (KW)},
Jennifer Y. H. Chan$^{1,2,6}$,
Kinwah Wu$^{1,7}$,
Curtis J. Saxton$^{4,5}$ and 
\newauthor Lidia van Driel-Gesztelyi$^{1,8,9}$
\\
$^{1}$Mullard Space Science Laboratory, University College London, Holmbury St Mary, Surrey, RH5 6NT, UK\\
$^{2}$Department of Physics and Astronomy, University College London, Gower Street, London, WC1E 6BT, UK\\
$^{3}$Institute of Astronomy, Department of Physics, National Tsing Hua University, Hsinchu 30013, Taiwan (ROC)\\
$^{4}$Physics Department, Technion -- Israel Institute of Technology, Haifa 32000, Israel\\
$^{5}$Department of Applied Mathematics, University of Leeds, Leeds LS2 9JT, UK\\ 
$^{6}$Lady Margaret Hall, University of Oxford, Norham Gardens, Oxford, OX2 6QA, UK\\ 
$^{7}$Perimeter Institute, 
  31 Caroline St.~N., Waterloo, Ontario N2L 2Y5, Canada\\
$^{8}$LESIA, Observatoire de Paris, Universit\'e PSL, CNRS, Sorbonne Universit\'e, Universit\'e de Paris, 5 place Jules Janssen,\\
92195 Meudon, France \\ 
$^{9}$Konkoly Observatory, MTA CSFK, H-1121 Budapest, Konkoly Thege M. \'ut 15-17, Hungary\\
}

\date{Accepted XXX. Received YYY; in original form ZZZ}

\pubyear{2019}

\begin{document}
\label{firstpage}
\pagerange{\pageref{firstpage}--\pageref{lastpage}}
\maketitle

\begin{abstract}
Faraday rotation measure at radio wavelengths is commonly used 
 to diagnose large-scale magnetic fields. 
It is argued 
  that the length-scales on which 
  magnetic fields vary in large-scale 
  diffuse astrophysical media 
  can be inferred from correlations in the observed RM. 
RM is a variable which can be derived 
  from the polarised radiative transfer equations 
  in restrictive conditions. 
This paper assesses the usage of 
  RMF (rotation measure fluctuation) analyses
  for magnetic field diagnostics 
  in the framework of polarised radiative transfer.  
We use models of various
   magnetic field configurations 
   and electron density distributions 
   to show how density fluctuations could affect 
   the correlation length of the magnetic fields 
   inferred from the conventional RMF analyses.
We caution against interpretations of RMF analyses 
   when a characteristic density is ill defined, 
   e.g. in cases of log-normal distributed 
   and fractal-like density structures. 
As the spatial correlations are generally not the same 
   in the line-of-sight longitudinal direction 
   and the sky plane direction, 
   one also needs to clarify 
   the context of RMF when inferring from observational data.  
In complex situations, 
   a covariant polarised radiative transfer calculation 
   is essential to capture  
   all aspects of radiative and transport processes, 
   which would otherwise ambiguate the interpretations 
   of magnetism in galaxy clusters 
   and larger-scale cosmological structures.  
\end{abstract}



\section{Introduction}
\label{sec-introduction}

Magnetic fields are present at all scales throughout the Universe, 
  from stars and substellar objects to galaxies, groups, clusters and large-scale structures such as filaments and voids
  \citep[see e.g.][for reviews]{Widrow2002,Widrow2012}.   
Stellar magnetic fields can be determined spectroscopically, e.g.\  
   by measuring Zeeman splitting in the optical spectral lines 
   for low-mass solar-like stars and magnetic white dwarfs 
\cite[e.g.][]{Wickramasinghe2000,Reiners2013},  
  from separations or locations of the cyclotron harmonic features in the optical/infrared spectra for accreting white dwarfs \cite[e.g.][]{Wickramasinghe1985,Wu1990}, 
  and from the X-ray spectra of neutron stars 
\cite[e.g.][]{Nagase1991,Santangelo1999,Staubert2019}.
Determination of magnetic field properties in larger astrophysical systems is less direct. 
For magnetic fields in diffuse astrophysical systems, 
  such as the interstellar medium (ISM), intracluster medium (ICM) and  intergalactic medium (IGM), 
  their properties are often inferred from the polarised radiation traversing and/or emitted from the media. 
Faraday rotation measure (RM)\footnote{%
`Faraday depth' and `rotation measure' can only be used interchangeably in the case of a single point source along the line-of-sight.}
has been identified as a diagnostic tool  
  for magnetic fields in our Galaxy
  \citep[see e.g.][]{Simard1980,Han1999,Brown2003, Gaensler2004,Brown2007,Haverkorn2008,Oppermann2012, Han2015, Han2017}, 
  nearby galaxies \citep[e.g.][]{Gaensler2005,Beck2009,Mao2010,Mao2017}
  and also some galaxy clusters \citep[e.g.][]{Carilli2002,Vogt2003,Clarke2004,Govoni2004,Brentjens2005,Bonafede2010,Kuchar2011,Vacca2018}. 

Recently, 
  there have also been studies 
  utilising the Faraday rotation of
  distant polarised radio sources 
  such as quasars
  \citep[e.g.][]{Kronberg2008, Xu2014redshift}
  and Fast Radio Bursts (FRBs)  
  \cite[e.g.][]{Xu2014b, Zheng2014,Akahori2016,Ravi2016,Vazza2018b,Hackstein2019},
  as a means to detect and probe cosmological magnetic fields. 
 These fields, permeating the cosmic web of filaments and voids, are weak,
  and their properties are often inferred statistically
  \citep[e.g.][]{Akahori2014b, Vernstrom2019}, 
  or indirectly constrained through the non-detection of GeV gamma-rays \citep[e.g.][]{Neronov2010, Taylor2011,Dermer2011, Takahashi2013}. 

The statistical characterisation of cosmological magnetic fields can be improved with a denser all-sky RM grid 
    from the Square Kilometre Array (SKA), including its pathfinders, 
    the Low Frequency Array (LOFAR),
    the Murchison Widefield Array (MWA),
    the Expanded Very Large Array (EVLA), 
    and its precursors, 
    the Australian SKA Pathfinder (ASKAP)
    and MeerKAT
    \citep[see e.g.][]{Gaensler2010, Beck2015, Johnston-Hollitt2015}.
How to properly characterise magnetic fields 
 beyond the scale of galaxy clusters is a challenge 
   in theoretical and observational astrophysics. 

Faraday Rotation Measure Fluctuation (RMF) analysis is proposed
  as a means to probe the structures of large-scale magnetic fields  \cite[e.g.][]{Akahori2010, Beck2013b}.
RM and RMF analyses 
  are essentially based on the theory of polarised radiative transfer
  under certain restricted conditions.
It is therefore important 
  to have a proper understanding of 
  the information we extract from the analyses 
  and under what conditions the analyses enable unambiguous interpretations.  
 
In this paper, 
  we examine the RMF analyses 
  in the context of polarised radiative transfer. 
We clarify 
the conditions 
  under which the RMF method 
  will give meaningful inferences 
  and identify the circumstances 
  where we should be cautious 
  when applying the method. 
We organise the paper as follows.
In \S\,\ref{sec-PRT}, we present 
  the formal covariant polarised radiative transfer formulation 
  and show how it reduces to the standard RM 
  under certain conditions.
In \S\,\ref{sec-RMF}, 
  we examine the RMF analysis 
  in the context of polarised radiative transfer. 
We also identify the mathematical and statistical properties of 
  the analyses. 
In \S\,\ref{sec-RM_var}, 
  we construct model density and magnetic field structures  
  and use them to test
  the validity of the RMF analyses.
We also discuss their astrophysical implications.
In \S\,\ref{sec-conclusions}, we present our findings and warnings.
Unless otherwise stated, 
  this work uses c.g.s.\ Gaussian units.

\section{Polarised radiative transfer}
\label{sec-PRT}

\subsection{Covariant transport in Stokes-parameter representation}%
\label{subsec-CPRT}

Under the conservation of photon number 
    and the conservation of phase-space volume 
    \citep[see][]{Fuerst2004, Younsi2012}, 
the covariant polarised radiative transfer equation may be expressed as 
\begin{equation}
\frac{{\rm d} \mathcal{I}_i }{\d \lambda_{\rm a}}\ \;\! 
   = \ \;\! -k_{\alpha}u^{\alpha}\; \Big\vert_{\lambda_{\rm a, co}} 
\left\{-\kappa_{ij,{\rm co}}\, \left(\frac{I_{j}}{\nu_{\rm co}^{3}}\right)+
\frac{\epsilon_{i,{\rm co}}}{\nu_{\rm co}^{3}}  \right\} 
\label{eq:covarPRT123}
\end{equation}
\citep[see][]{Chan2019}.
Here,
$\nu$ is the radiation frequency,
$\mathcal{I}_i$ is the Lorentz-invariant Stokes vector, 
$\lambda_{\rm a}$ is the affine parameter and
$-k_{\alpha}u^{\alpha}\; \big\vert_{\lambda_{\rm a, co}}$ 
 is the projection factor for a photon with a 4-momentum $k^{\alpha}$ 
  travelling in a fluid with a 4-velocity $u^{\beta}$.
The subscript `co' denotes 
   that the quantity is evaluated in the reference frame 
   co-moving with the fluid. 
The transfer matrix, $\kappa_{ij,{\rm co}}$, accounts for the absorption and Faraday propagation effects, 
    while the emission coefficients are defined by $\epsilon_{i,{\rm co}}$.

In a Friedmann-Robertson-Walker (FRW) universe,
  the displacement $s$ as a function of redshift $z$ is given by
\begin{equation} 
\frac{\d s}{\d z} 
= 
\frac{c}{H_{0}}\,({1+z})^{-1}
   \left[\Omega_{{\rm r},0}(1+z)^{4}+\Omega_{{\rm m},0}(1+z)^{3}+\Omega_{\Lambda,0}\right]^{-\frac{1}{2}}  
\label{eq:dldz}
\end{equation}
\citep[see e.g.\,][]{Peacock1999}, 
where  
 $H_{0}$ is the Hubble parameter, $\Omega_{{\rm r},0}$, $\Omega_{{\rm m},0}$ and $\Omega_{\Lambda,0}$ are the dimensionless energy densities of relativistic matter and radiation, non-relativistic matter, 
and a cosmological constant (dark energy with an equation of state of $w \equiv -1$), respectively. 
The subscript ``0" denotes that the quantities are measured at present\,(i.e. $z=0$).  
As such, equation~(\ref{eq:covarPRT123}) becomes
\begin{equation}
\frac{\rm d}{\d z}
\begin{bmatrix} 
{\mathcal I}\\
{\mathcal Q}\\
{\mathcal U}\\
{\mathcal V}
\end{bmatrix} 
=   
 ({1+z})
 \left\{
 -    \begin{bmatrix} \kappa & q & u & v \\ q &\kappa & f & -g \\ u & -f & \kappa&h  \\ v& g & -h& \kappa  \end{bmatrix} 
\begin{bmatrix} 
{\mathcal I}\\
{\mathcal Q}\\
{\mathcal U}\\
{\mathcal V}
\end{bmatrix} 
 +\begin{bmatrix}
 \epsilon\subI \\ \epsilon\subQ \\ \epsilon\subU \\ \epsilon\subV
 \end{bmatrix} 
\frac{1}{\nu^3} \right\} 
\frac{\d s}{\d z}    
\label{eq:covarPRTinz}
\end{equation} 
\citep{Chan2019}, where 
    $\kappa,\, q,\, u,\, v$ are the absorption coefficients,
    $\epsilon$ are the emission coefficients,  
    $f$ is the Faraday rotation coefficient, and 
    $g$ and $h$ are the Faraday conversion coefficients. 
The invariant Stokes parameters are related to the usual Stokes parameters by 
$[\,{\mathcal I}\, {\mathcal Q}\, {\mathcal U}\, {\mathcal V}\,]^{\rm T}  = [\,I\, Q\, U\, V\,]^{\rm T} /\nu^{3}$\,.

In a local frame, the covariant polarised radiative transfer equation in (\ref{eq:covarPRTinz}) reduces to the standard polarised radiative transfer equation: 
\begin{equation} 
  \frac{\rm d}{{\rm d} s}   
   \begin{bmatrix} I \\ Q \\ U \\ V \end{bmatrix} 
   =  
    -    \begin{bmatrix} \kappa & q & u & v \\ q &\kappa & f & -g \\ u & -f & \kappa&h  \\ v& g & -h& \kappa  \end{bmatrix} 
         \begin{bmatrix} I \\ Q \\ U \\ V \end{bmatrix} 
        +   \begin{bmatrix}
        \epsilon\subI \\ \epsilon\subQ \\ \epsilon\subU \\ \epsilon\subV 
        \end{bmatrix}   \  .
\label{eq-PRT}
\end{equation}
The Stokes parameters $[\,I,\, Q,\, U,\, V\,]$ are observables, 
and their combination gives rise to different derived quantities, including the total degree of polarisation 
$\Pi_{\rm tot} = \sqrt{Q^2 + U^2 + V^2}/I\, (\leq1)$, 
    the degree of linear polarisation 
    $\Pi_{\rm l} =   \sqrt{Q^2 + U^2}/I$,  
      the degree of circular polarisation 
      $\Pi_{\rm c} = V/I$, 
      and the polarisation angle $\varphi= (1/2) \tan^{-1}(U/Q)$
      \citep[see e.g.][]{Rybicki1979}.

\subsection{Derivation of rotation measure}%
\label{subsec-RM}  

In the absence of absorption and emission, we can set
$\kappa = q = u = v =0$ and
$[\, \epsilon\subI,\, \epsilon\subQ,\, \epsilon\subU,\, \epsilon\subV\,] = 0$,
therefore imposing
    ${\rm d}I /{\rm d}s = 0$, and
 \begin{equation} 
  \frac{\rm d}{{\rm d} s}   \begin{bmatrix}  Q \\ U \\ V \end{bmatrix}   =  
    -    \begin{bmatrix}    0 & f & -g \\   -f & 0 &h  \\ g & -h& 0  \end{bmatrix} 
         \begin{bmatrix}  Q \\ U \\ V \end{bmatrix}    \  . 
\label{eq-PRT_x}
\end{equation}
In situations where
the circular polarisation is insignificant 
  and the conversion between linear and circular polarisation is negligible,  
  we may consider only two linearly polarised Stokes components
  in the polarised radiative transfer calculation. 
The polarised radiative transfer equation then takes a simplified form: 
\begin{equation} 
  \frac{\rm d}{{\rm d} s}   \begin{bmatrix}  Q \\ U  \end{bmatrix}   =  
    -    \begin{bmatrix}    0 & f \\   -f & 0   \end{bmatrix} 
         \begin{bmatrix}  Q \\ U  \end{bmatrix}    \  .    
\label{eq-PRT_restrictive}
\end{equation}  
The Faraday rotation coefficient $f$ is the sole parameter in this equation. 
It is determined by the properties of free electrons and the magnetic field along the line-of-sight.

An astrophysical plasma 
  may contain both thermal and non-thermal electrons. 
If the fraction of non-thermal electrons is small, 
  the 
  conversion between the two linearly polarised Stokes components  
  is determined mainly by the thermal electrons. 
With only thermal electrons present, 
  the Faraday rotation coefficient is
\begin{equation} 
  f_{\rm th}  =  \frac{\omega_{\rm p}^2 \cos \theta}{ c\, \omega_{\rm B}} 
    \left(\frac{\omega_{\rm B}^2}{\omega^2 - \omega_{\rm B}^2}  \right)   
\end{equation}  
  \citep{Pacholczyk1977}, where 
  $\omega = 2 \uppi \nu$ is the angular frequency of radiation,
  $\omega_{\rm p} = (4\uppi n_{\rm e,th} e^2/m_{\rm e})^{1/2}$ is the plasma frequency, 
  $\omega_{\rm B} = (e B/ m_{\rm e}c)$ is the electron gyro-frequency, 
  $n_{\rm e,th}$ is the thermal electron number density,
  $B$ is the magnetic field strength
  and $\theta$ is the angle between the magnetic field vector and the line-of-sight. 
Here, $c$ is the speed of light, $e$ is the electron charge, 
  and $m_{\rm e}$ is the electron mass. 
In the high-frequency limit (i.e. $\omega \gg \omega_{\rm B}$),
  the Faraday rotation due to only thermal electrons can be expressed as, 
\begin{equation}  
  f_{\rm th}  =  
  \frac{1}{\uppi }\,
  \left(\frac{e^3}{m_{\rm e}^2c^4} \right) \,
     n_{\rm e,th}\, {B}_\parallel \, \lambda^2  \ , 
\end{equation}    
    where  
    ${B}_{\parallel} = |\boldsymbol{B}| \cos \theta$ 
    is the magnetic field along the line-of-sight and    
    $\lambda  = 2\uppi c /\omega$ is the wavelength of radiation.
The corresponding expression for Faraday rotation due to only non-thermal electrons is
\begin{equation}  
  f_{\rm nt}  =  \frac{1}{\uppi } \left(\frac{e^3}{m_{\rm e}^2c^4} \right)  \zeta (p, \gamma_{\rm i}) \, 
     n_{\rm e,nt} \, {B}_\parallel \,  \lambda^2  \ ,    
\end{equation} 
    where the factor, 
\begin{equation} 
   \zeta (p, \gamma_{\rm i})  =  \frac{(p-1)(p+2)}{(p+1)}  \left( \frac{\ln \gamma_{\rm i}}{{\gamma_{\rm i}}^2}  \right) \ ,
\end{equation} 
for $p >1$, 
    assuming an isotropic distribution of non-thermal electrons with a power-law energy spectrum of index $p$
    \citep{Jones1977a}.
The number density of non-thermal electrons is $n_{\rm e,nt}$,
  and
  $\gamma_{\rm i}$ is their low-energy cut-off.  

In a plasma consisting of thermal electrons 
  plus non-thermal electrons,  
  the relative strength of their contributions to the Faraday rotation is therefore
\begin{equation} 
   \frac{f_{\rm nt}}{f_{\rm th}}
     \approx    \zeta(p, \gamma_{\rm i})
     \left( \frac{n_{\rm e, nt}}{n_{\rm e, th}}\right) \ ,
     \label{eq.f.ratio}
\end{equation} 
   provided that neither $n_{\rm e,nt}$ nor $n_{\rm e,th}$ correlates or anti-correlates significantly 
   with ${B}_\parallel$\footnote{%
   A similar relation was given in \cite{Jones1977a} 
   for the 
   relative contributions of relativistic and thermal electrons to the Faraday rotation.
Their relation is expressed in terms of the spectral index $\alpha$ 
   of the optically thin power-law synchrotron spectrum. 
The relation (\ref{eq.f.ratio}) here 
   is expressed in terms of the power-law index $p$ 
   of the electron energy distribution, 
   which is intrinsic to the magneto-ionic medium. 
   Note that
   $\alpha = (p - 1)/2$. }. 

From the restrictive polarised radiative transfer equation~(\ref{eq-PRT_restrictive}) which only has two linear Stokes components, 
   it can easily be shown
   that the change in the linear polarisation angle 
   along the line-of-sight is 
\begin{equation} 
  \frac{{\rm d}\varphi}{{\rm d} s}  =  \frac{1}{2}\left(\frac{1}{U^2+Q^2}\right)
     \left(Q \frac{{\rm d}U}{{\rm d} s}-U\frac{{\rm d}Q}{{\rm d} s} \right)   = \frac{f}{2} \ . 
\label{eq-dphids}
\end{equation} 
With only thermal electrons in a sufficiently weak magnetic field where  $\omega_{\rm B} \ll \omega$, 
  a direct integration of equation~({\ref{eq-dphids}}) with $f = f_{\rm th}$ yields 
\begin{equation} 
  \varphi(s)    =    \varphi_0 + \frac{2\uppi e^3}{m_{\rm e}^2(c\, \omega)^2} 
     \int^s_{s_0} {\rm d}s' \, n_{\rm e, th}(s') \, B_\parallel(s') \ . 
\end{equation} 

Rotation measure (RM) is defined as 
\begin{equation} 
{\cal R}  =  (\Delta \varphi) \;\! \lambda^{-2} =   \left(\varphi - \varphi_0\right)\;\! \lambda^{-2} \ . 
\end{equation}    
The polarised radiative transfer equations~(\ref{eq-PRT}), (\ref{eq-PRT_x}) and (\ref{eq-PRT_restrictive}) 
  are linear, 
  thus the contributions to the Faraday rotation coefficient by a collection of thermal and non-thermal electrons are additive. 
The RM for radiation traversing a magnetised plasma between an interval $s_0$ and $s$
  is therefore 
\begin{equation} 
 {\cal R}(s) =  \frac{e^3}{2\uppi m_{\rm e}^2c^4}   \int^s_{s_0} {\rm d}s' \ n_{\rm e}(s') \, \Theta (s') \, {B}_\parallel(s')  \ , 
\label{eq-R_x}
\end{equation}  
  where
  $n_{\rm e}$ is the total electron number density, and
    $\Theta(s) = 1 - \Upsilon(s) \, [1- \zeta(p, \gamma_{\rm i})]\big\vert_{s}$ is the weighting factor of $n_{\rm e}$ contributing to the Faraday rotation effect, accounting for both thermal and non-thermal electron populations, with $\Upsilon(s)$ the local fraction of non-thermal electrons.
If only thermal electrons are present, $\Upsilon(s)=0$ such that $\Theta(s) = 1$,
   hence recovering the widely-used formula in RM analysis
   of magnetised astrophysical media
   \citep[see e.g.][]{Carilli2002}:
\begin{equation} 
 {\cal R}(s)  =   0.812  \int^s_{s_0} \frac{{\rm d}s'}{{\rm pc}} \!
      \left(\frac{n_{\rm e, th}(s')}{{\rm cm}^{-3}} \right)\! \left( \frac{{B}_\parallel(s')}{\mu{\rm G}}\right) \,  {\rm rad}\;\!{\rm m}^{-2} \  . 
\label{eq:RM_int}
\end{equation}

\section{Rotation Measure Fluctuations} 
\label{sec-RMF}

\subsection{Computing rotation measure in a discrete lattice} 
\label{subsec-RM-comp}
Practical calculations of polarised radiative transfer in an inhomogeneous medium often require sampling the medium into discrete segments 
  that have small internal variations in physical properties. 
Suppose we divide the radiation propagation path length $L$ 
   into $N$ intervals of lengths $\Delta s$, i.e.\ $L = \sum_{i=1}^N \Delta s(i)$. 
Then the integral in equation~(\ref{eq-R_x}) can be approximated 
  by summing contributions from all segments
\begin{equation} 
  {\cal R}(s)   =   \frac{e^3}{2\uppi m_{\rm e}^2c^4}  
   \sum_{i=1}^{N} {\Delta s}(i) \ n_{\rm e}(i) \, \Theta (i) \, B_\parallel(i)  \ ,   
\end{equation} 
  where
  $n_{\rm e}$, 
  $\Theta$ and 
  $B_\parallel$ are 
  evaluated at the centre of each interval, $s_i$. 
If the magnetic fields 
     have uniform strengths and unbiased random orientations, then
    $B_\parallel$ will have a symmetric probability distribution: 
    $P (B_\parallel) = P (-B_\parallel)$.
With $\Delta s > 0$, $n_{\rm e} \geqslant 0$  and $\Theta \in [\;\!0,1\;\!]$,  
   the symmetry in the probability distribution of $B_\parallel$ 
   implies that the expectation value of RM
\begin{equation}  
  \langle  {\cal R}  \rangle   =   \frac{e^3}{2\uppi m_{\rm e}^2c^4} 
      \sum_{i=1}^{N}  \langle {\Delta s}\   {n_{\rm e}}\,  \Theta\,  B_\parallel \rangle\,\big\vert_i \ \ = \ \ 0 \   , 
\label{eq-expR}
\end{equation} 
  where $\langle ... \rangle$ denotes the ensemble average of the variables.  

Supposing that   
    $n_{\rm e}$, $\Theta$ and $B_\parallel$ are 
    incoherent
    among the intervals $\Delta s$,  
then $n_{\rm e}$, $\Theta$ and $B_\parallel$ 
    are the only independent variables 
    for computing the RM of a cell  
    defined by an interval.
Moreover, if the medium does not evolve during the radiation's propagation,   
  $n_{\rm e}$, $\Theta$, $B_\parallel$ and their products 
  are also exchangeable variables. 
Under the ergodic condition, the ensemble averages of independent and exchangeable variables 
  can be replaced by the averages of over the path length, i.e.\   for a sufficiently large $N$,  
\begin{equation} 
\langle X \rangle  =   \frac{1}{N} \sum_{j=1}^N X(s_j) \ \ = \ \ \langle X(s_j) \rangle_s \ .     
\end{equation}   
Thus,  
\begin{equation}  
\langle  {\cal R}  \rangle   =   \frac{e^3}{2\uppi m_{\rm e}^2c^4} \, 
  N \langle {\Delta s}\   {n_{\rm e}}\,  \Theta\,  B_\parallel \rangle_s \ \ = \ \ 0 \ .      
\label{eq-meanR}
\end{equation}   
Moreover, if
${\boldsymbol B}$, 
  $n_{\rm e}$ and 
  $\Theta$
  do not correlate with each other, we have 
\begin{equation}  
  \langle  {\cal R}  \rangle  =   \frac{e^3}{2\uppi m_{\rm e}^2c^4} \,
  N\;\! \langle \Delta s\rangle_s  \langle {n_{\rm e}}\rangle_s \langle{\Theta}\rangle_s \langle B_\parallel \rangle_s 
    \ \ = \ \ 0 \ . 
\end{equation}

\subsection{Rotation measure fluctuations as a restrictive 
  autoregression (AR) process}  
\label{subsec-RMF_AR}

Note that an observable ${\tilde {\cal O}}_k$ on the lattice grid $k$ 
  in an AR(1) (autoregression of order one) process on a 1-D lattice  
  is given by the recursive relation:     
\begin{align}
  {\tilde {\cal O}}_k & =   \rho\ {\tilde {\cal O}}_{k-1} + \varepsilon_k  \nonumber \\ 
                             & =   \rho \left(  \rho\ {\tilde {\cal O}}_{k-2}  + \varepsilon_{k-1}    \right) 
                             + \varepsilon_k \nonumber \\ 
      & \cdot  \cdot   \cdot   \cdot    \cdot  \cdot \nonumber   \\   
          & =  \rho^m \ {\tilde {\cal O}}_{k-m} + \sum_{j=0}^{m-1} \rho^{j} \ {\varepsilon_{k-j}}  
\label{eq-AR1}
\end{align}        
   \citep[see e.g.][]{Box1976, Anderson1976, Grunwald1996}, 
   where $\rho$ is a parameter, 
   and $\varepsilon_k$ is an iid (independent, identically distributed) variable 
   with an expectation value ${\rm E} (\varepsilon_k) = \langle \varepsilon_k \rangle = 0$ 
   and a variance ${\rm Var}(\varepsilon_k)  = \left[ \sigma(\varepsilon_k) \right]^2$.     
For a finite or semi-infinite lattice, which is truncated at $j=0$, 
  at which the observable ${\tilde {\cal O}}_0$ is well defined, we can rewrite  equation~(\ref{eq-AR1}) as    
\begin{equation}       
    {\tilde {\cal O}}_k  =  \rho^k \ {\tilde {\cal O}}_0 + \sum_{j=1}^k \rho^{k-j} \ {\varepsilon_j}  \ . 
    \label{eq-AR1-finitelattice}
\end{equation}   
  
For a polarised radiation's propagation path consisting of $N$ segments 
  with approximately coherent Faraday rotation properties,    
  the polarisation angle at the end of the $k^\mathrm{th}$ segment is given by 
\begin{align}
  \varphi_k & =   \varphi_{k-1} + \Delta \varphi_k    \nonumber \\ 
                  & =   \varphi_{k-2} + \Delta \varphi_{k-1}  + \Delta \varphi_k    \nonumber \\  
        & \cdot  \cdot   \cdot   \cdot    \cdot  \cdot  \nonumber \\        & =  \varphi_0 + \sum_{j=1}^{k} \Delta \varphi_{j}   \   , 
\label{eq-Dphi}                   
\end{align} 
  where $\Delta \varphi_k$ is the rotation of the polarisation angle in the $k^\mathrm{th}$ segment, 
  and the polarisation angle measured by the observer is simply  $\varphi_N$. 
  Comparing equations~(\ref{eq-Dphi}) and (\ref{eq-AR1-finitelattice})  
   reveals that the evolution of the polarisation angle along the  radiation's propagation 
   is an AR(1) process with a constant parameter $\rho =1$,  
   provided that $\langle \Delta \varphi_j \rangle  = 0$ 
   and that ${\rm Var}(\Delta \varphi_j)$ is well defined and computable.  
An AR(1) process is a Markov process \citep[see][]{Anderson1976}, 
  and an AR(1) process with $\rho = 1$ is also known as a simple random-walk.

The rotation measure across the propagation path of the radiation is
   ${\cal R}\big\vert_{N} =  (  \varphi_N - \varphi_0)\;\!  \lambda^{-2}$. 
Hence, from equation~(\ref{eq-meanR}), we obtain   
\begin{equation}  
  \bigg\langle \sum_{j=1}^{N} \Delta \varphi_{j} \bigg\rangle_s  =  
  \lambda^2\ \langle {\cal R}\rangle_s   \ \  =  \ \  0   \  .    
\end{equation}    
As the expectation value and the variance of $(\varphi_N - \varphi_0)$ are 
\begin{align}
  {\rm E} \big[\,\varphi_N - \varphi_0 \, \big] & =     0 \  ; \\  
   {\rm Var}   \big[\, \varphi_N - \varphi_0 \, \big] & =    N \ \sigma^2  \ ,  
\end{align} 
  respectively, with $\sigma^2 = \big\langle {\Delta \varphi_j}^2 \big\rangle$, the standard deviation of ${\mathcal R}$ in the radiation's propagation direction 
  is therefore  
\begin{align}  
  \sigma_{\cal R} 
    & = \sqrt{N}\ 
    \left [ \big\langle {\Delta \varphi_j}^2 \big\rangle \right]^{1/2} \;\! \lambda^{-2}  \nonumber \\
    & =  \frac{e^3\,  \sqrt{N}}{2\uppi m_{\rm e}^2c^4}  \ 
    \left[ \big\langle {\Delta s}^2\,  n_{\rm e}^2\, \Theta^2\, {B_\parallel}^2 \big\rangle \right]^{1/2}  \nonumber \\ 
      & =  \frac{e^3\,  \sqrt{N}}{2\uppi m_{\rm e}^2c^4}  \ 
    \left[ \big\langle {\Delta s}^2\,  n_{\rm e}^2\, \Theta^2\, {B_\parallel}^2 \big\rangle_s \right]^{1/2}     \ .   
    \label{eq:sigmaRM}
\end{align}  
Note that the rotation measure fluctuation along a radiation propagation path consisting of coherent segments is  proportional to the square root of the number of the segments ($\sqrt{N} = \sqrt {L/ \langle \Delta s \rangle_s}$),   
  a characteristic of a simple random-walk process, 
  where the root mean square displacement is proportional to the square root of the number of steps. 
Here, the root mean square of properties within a step size is  
   $\left[ \big\langle {\Delta s}^2\,  n_{\rm e}^2\, \Theta^2\, {B_\parallel}^2 \big\rangle_s \right]^{1/2}$.  
In the specific condition that the interval segments have equal length, $\overline{\Delta s}$, 
   and $(n_{\rm e}\,\Theta)$ 
   does not vary along the line-of-sight,  
equation~(\ref{eq:sigmaRM}) becomes 
\begin{equation}  
  \sigma_{\cal R}   =  \frac{e^3}{2\uppi m_{\rm e}^2c^4}  \, 
      \sqrt{\frac{L}{\overline {\Delta s}}}\ {\overline {\Delta s}} \ n_{\rm e}\, \Theta 
    \left[ \big\langle {B_\parallel}^2 \big\rangle_s \right]^{1/2}     \ .  
    \label{eq-sigRM_ana}
\end{equation}     

A similar but more rigorous expression can be obtained 
  if there is no correlation between 
  electron number density and the magnetic fields. 
In this case,  
equation~(\ref{eq:sigmaRM}) becomes 
\begin{equation}  
  \sigma_{\cal R}   =  \frac{e^3}{2\uppi m_{\rm e}^2c^4}  \, 
      \sqrt{\frac{L}{\overline {\Delta s}}}\ {\overline {\Delta s}} \ {\overline{n_{\rm e} \, \Theta }}
    \left[ \big\langle {B_\parallel}^2 \big\rangle_s \right]^{1/2}  
    \     , 
    \label{eq-sigRM_ana-mean}
\end{equation}  
 with ${\overline{n_{\rm e}\, \Theta }}$ 
   denoting the mean value of $({n_{\rm e}}\, \Theta)$. 
Additionally, in the presence of only thermal electrons, 
 then $\Theta =1$ uniformly, and   
\begin{align}
\sigma_{\cal R} & =   
\frac{e^3}{2\uppi m_{\rm e}^2c^4} \,
\sqrt{\frac{L}{\overline {\Delta s}}}\ 
{\overline {\Delta s}} \,
    \overline{n}_{\rm e, th} \,
    B_{\parallel {\rm rms}} 
     \nonumber \\ 
 & = 0.812 \, 
 \sqrt{\frac{L}{\overline {\Delta s}}}\, 
  \left(\frac{ {\overline {\Delta s}}}{\rm pc} \right)\,
  \left( \frac{ \overline{n}_{\rm e, th}}{{\rm cm}^{-3}} \right) \,
  \left( \frac{ B_{\parallel {\rm rms}}}{\mu{\rm G}} \right)
    \ 
    {\rm rad}~{\rm m}^{-2}  \ .
\label{eq-sigrm_choryu}
\end{align}

Most observational or numerical studies
   use either one of the expressions given in equations 
   (\ref{eq-sigRM_ana-mean}), and  (\ref{eq-sigrm_choryu}) 
    in their RM fluctuation analysis.
These include investigations of magnetic fields in galaxy clusters   
  or in large-scale structures
  \citep[e.g.][]{Sokoloff1998, Blasi1999, 
  Dolag2001, Govoni2004, 
   Subramanian2006, Cho2009, Sur2019}.  
Note that 
  the two expressions above are 
  not always explicitly distinguished   
  in studies of RM fluctuations. 
The $\sigma_{\mathcal R}$ derivations from equation (28) to equations (29), (30) and (31)
   rely on subtly different assumptions regarding
  the electron density spatial distributions
  and their relation or correlation with the magnetic fields.
For instance,
   it matters whether local quantities are multiplied before spatial averaging,
   or averaged separately then multiplied.
Note also that, 
  in reality,  
  the condition of constant electron number density,  
  or/and the condition of electron number density and magnetic field 
  being uncorrelated, are generally not satisfied. 
We should therefore bear in mind 
  which underlying assumptions have been used,  
  and they should be stated explicitly 
  when interpreting the magnetic field structures 
  using the observed RM statistics. 
Furthermore, while $\sigma_{\mathcal R}$ is observed on the sky plane,
  it is calculated over the radiation's propagation path,
  with the application of a random walk model along the line-of-sight
  and invoking other explicit assumptions we made above.

\subsection{Fluctuations of density and magnetic fields in parallel and in perpendicular directions}
\label{subsec-pp_flucation}

\begin{figure}
    \centering
 \includegraphics[width=0.47\textwidth]{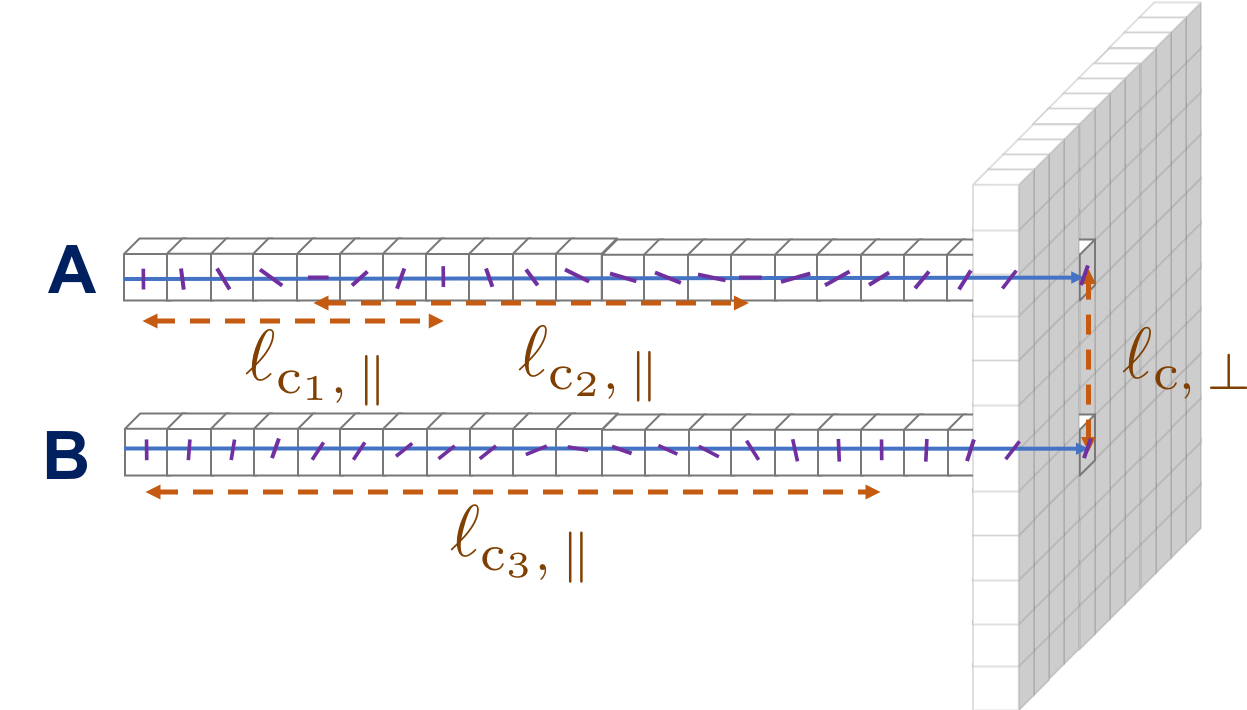}
      \caption{Illustration of a case where $\sigma_{{x},\,{\perp}} \neq \sigma_{{x},\,{\parallel}}$, 
  emphasising that polarisations are modified along the line-of-sight, 
  but what we observe is only the resulting angle on the orthogonal image plane. 
  In this demonstration, line-of-sight {\bf A} exhibits 
  two characteristic lengths of polarisation fluctuation, 
  $\ell_{{\rm c}_{1},\,\parallel}$ and $\ell_{{\rm c}_{2},\,\parallel}$,
  as illustrated by the number of cells spanning a full rotation,
  while for line-of-sight {\bf B},  $\ell_{{\rm c}_{3},\,\parallel}$.
  None of these coincides with $\ell_{{\rm c},\,\perp}$ on the image plane.}
    \label{fig:para_perp_demo}
\end{figure}

The polarisation of radiation 
  at a location on the sky plane, and hence the celestial sphere, 
  is determined by the magneto-ionic properties of plasma
  along the line-of-sight 
  (specified by the propagation unit vector ${\boldsymbol {\hat k}}$). 
  On cosmological scales, 
  the transfer of radiation along the line-of-sight 
  is the transfer of radiation from the past to the present. 
  Consequently, 
the statistical properties 
  of the observed polarisation signatures across a sky plane 
  depend on two factors: 
  (i) the spatial variations of the magneto-ionic plasma properties 
    at different cosmological epochs, and  
  (ii) the temporal variations of the magneto-ionic plasma properties 
    as the Universe evolved.  
Note that these two factors are not always mutually independent. 
It is their convolution that will determine the variations of the observable variables 
  along the ray as it propagates 
  (i.e.~in ${\boldsymbol {\hat k}}$ direction, denoted by $\parallel$) 
  and the variations of the observable variables 
  among the collection of rays reaching the sky plane (i.e.~ in directions $\perp$ to ${\boldsymbol {\hat k}}$). 
More importantly,
  there is no guarantee that 
  these two types of fluctuations
  are statistically identical. 
In other words, 
 if we use a simple representation 
  with two independent orthogonal components, 
  designated to be parallel and perpendicular to ${\boldsymbol {\hat k}}$, 
 we cannot simply assume that 
  $\sigma_{{x},\,{\perp}} = \sigma_{{x},\,{\parallel}}$, 
  where the quantity $x \in \{\,Q,\,U,\,V,\,{\Delta \varphi},\  {\rm or}\ {\mathcal R}\, \}$. 
  In general, we have two separate correlation lengths, 
  $\ell_\parallel$ and $\ell_\perp$, for each plasma quantity, 
  e.g. the electron number density $n_{\rm e}$  
  (which is a scalar)  
  and the magnetic field ${\boldsymbol B}$  
  (which is a vector). See Fig.~\ref{fig:para_perp_demo} for an illustration.


The question now is whether we can take  
   a correlation length  
   derived from a polarisation signature across the sky plane 
   as the characteristic correlation length scale  
   over which the cosmological magnetic fields 
   vary spatially
   or, alternatively, over time.
Firstly, the polarised radiative transfer equation shows that 
  the rotation of the polarisation angle 
  only depends on the magnetic field parallel to the line-of-sight, 
  i.e. ${\boldsymbol B}_{\parallel}$.  
The perpendicular component 
  of the magnetic field ${\boldsymbol B}_{\perp}$ is irrelevant in this respect. 
Therefore, in each individual ray,
  the local polarisation fluctuations are only caused by
  the fluctuations of the parallel field component,  
  ${\boldsymbol B}_{\parallel}$,   
  and the fluctuations of the electron number density $n_{\rm e}$ 
  and energy distribution. 

Secondly, 
  the fluctuations of polarisation properties 
  along individual rays are not directly observable. 
Instead, observations reveal
  a ``polarisation map'' on the celestial sphere, 
  which represents the polarisation signatures 
  of a collection of end-points of the path-integrated polarised rays.  
If the rays are independent, 
  we would observe variations in the polarisation signatures, 
  such as the RM fluctuations, 
  even when the magneto-plasma 
  is statistically spatially uniform at any cosmological epoch. 
In this situation, 
  the observed RM fluctuations 
  reflect 
  the convolution of the fluctuations 
  in $\vert {\boldsymbol B}_\parallel \vert$ and $n_{\rm e}$ 
  along the line-of-sight, 
  i.e. not simply an effect arising from 
  the presence of spatial structures. 
Note that there are additional subtleties 
  in assessing the local variations of polarisation signature 
  along a ray. 
Suppose that the electron number density and its energy spectrum 
  are uniform in both space and time, 
  there still exists an ambiguity 
  in determining the fluctuation 
  of the magnetic field ${\boldsymbol B}$, 
  as the polarisation angle 
  rotates depending on value of 
  $\vert {\boldsymbol B}_\parallel \vert$, 
   which equals to $\vert {\boldsymbol B} \vert \cos \theta$, 
  at an unknown angle $\theta = \cos^{-1}  
    ({\boldsymbol {\hat k}} \cdot {\boldsymbol {\hat B}})$.       
Thus, 
   there are two aspects in the magnetic field fluctuations, 
   one concerning the field magnitudes, 
   and another concerning the field orientations. 
Magnitude fluctuations and orientation fluctuations   
   can arise from different processes. 
For instance,   
  the variations in the magnetic field orientation ${\boldsymbol {\hat B}}$ 
  may indicate the characteristic size 
    of the astrophysical system 
  or the magnetic sub-domain of the cosmic magneto-ionic plasma, 
  while  
   the variations in the magnetic field magnitude 
   $\vert {\boldsymbol B}\vert$ 
   along the line-of-sight 
   would inform us 
   about the changes in the global magnetic energy density 
   as the Universe evolves. 
 Fluctuations of $\boldsymbol {\hat B}$ and those of $\vert {\boldsymbol B} \vert$ 
   can arise from different mechanisms 
   and/or operate on different characteristic time scales.  
 Thus, $\sigma_{{\abs{\boldsymbol{B}}}}\big\vert_{\parallel,\,{\rm or}\,\perp} = \sigma_{\boldsymbol {\hat B}}\big\vert_{\parallel,\,{\rm or}\,\perp}$ 
  do not usually hold.


\begin{figure}
    \centering
    \includegraphics[width=0.47\textwidth]{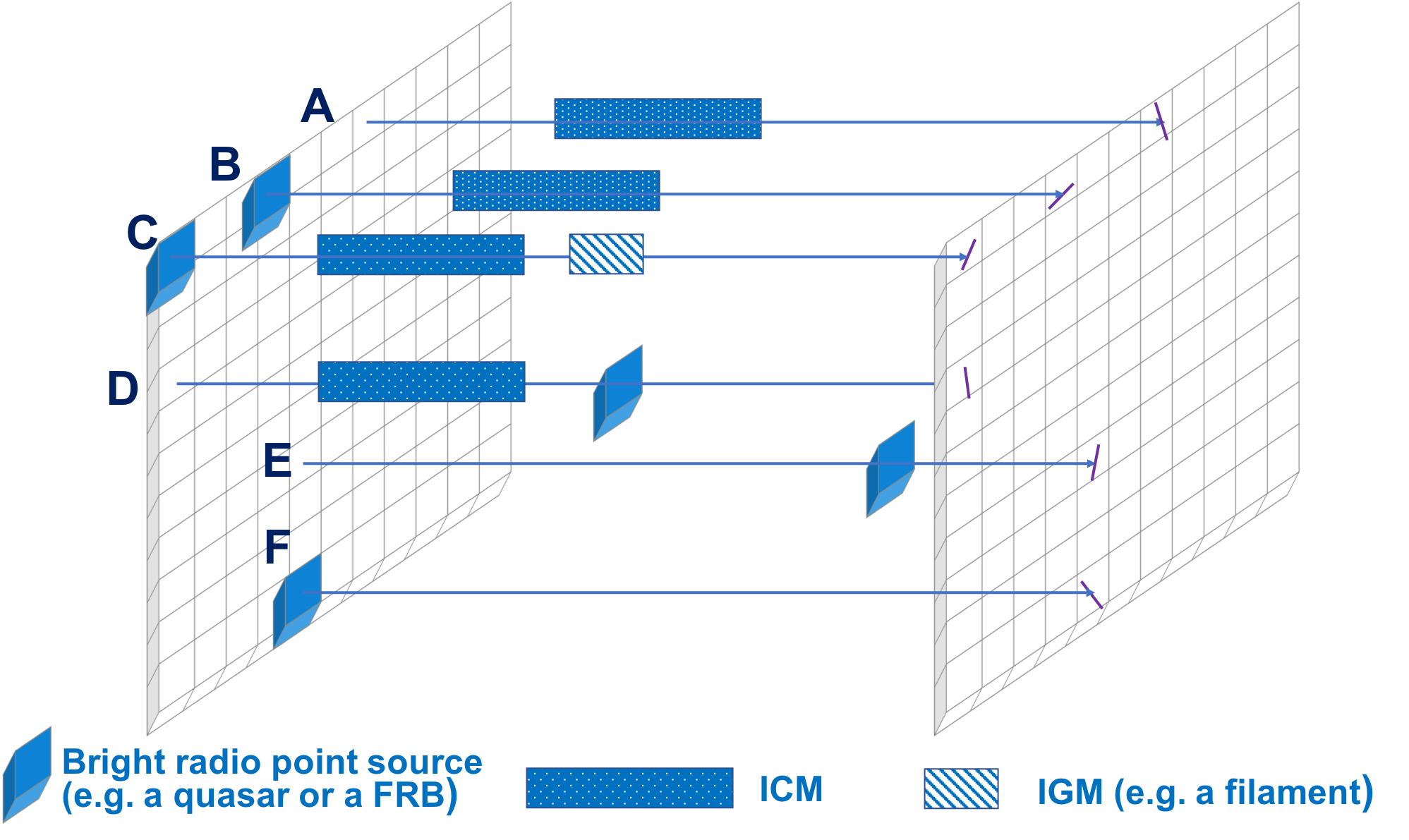}
      \caption{Illustration of how different astrophysical conditions give rise to different polarisation fluctuations due to e.g. 
      (i) the presence or absence of bright background sources (lines-of-sight {\bf A} and {\bf B}),
      (ii) the presence or absence of multiple sources with different Faraday depths (lines-of-sight {\bf B} and {\bf C}), 
      (iii) different positional orders of sources  (lines-of-sight {\bf B} and {\bf D}), and 
      (iv) change of radiation frequency due to the Universe's expansion, and/or the 
      presence of sources either at low or high redshift
      (lines-of-sight {\bf E} and {\bf F}).
      }
      \label{fig:los_demo}
\end{figure}
  
Fig.~\ref{fig:los_demo} illustrates some example scenarios  
  that give rise to different observational  polarisation signals%
  \footnote{
 Inferring the magneto-ionic properties of the line-of-sight sources and the intervening plasmas 
  from the polarised sky data, 
  which has a $(2+1)$-D structure, where the ``$+1$" corresponds to the time axis or cosmological redshift,
  is an inverse problem. 
 In forward theoretical modelling, 
   the polarisation signals are, however, 
   determined by the cosmological polarised radiative transfer, which is in a $(3+1)$-D format, where the line-of-sight direction also aligns with the axis of cosmological time.}. 
The types and number of sources, 
  and the magneto-ionic properties of the intervening plasmas 
  vary along each line-of-sight 
  and vary among the lines-of-sight
  across the sky. 
Magnetic fields are vectors, 
   and therefore 
   possess two structural traits:
   one in the field strength (or energy density), 
   and another in the field orientation. 
Both aspects are 
  essential to determine the 
  properties of all-sky polarisation.   
Faraday rotation also depends on 
   the line-of-sight electron number densities, 
   and 
   the electron number densities and the magnetic fields
   are usually interdependent.
Depending on what mechanisms 
  generate and/or amplify
  the magnetic fields, 
  local and non-local correlations 
  between the two quantities could occur. 
As such, 
  we cannot simply take the face values of 
  the correlation lengths obtained from 
   RM fluctuation analyses as they appear,   
   without careful consideration. 
In the analysis of cosmological-scale magnetism 
  we also need to consider
  effects due to cosmological expansion. 
Large-scale magnetic fields would evolve 
  with the cosmological structure 
  \citep[see e.g.][]{Dolag2005,Cho2009,Ryu2012,Barnes2012, Barnes2018MNRAS,Marinacci2015,Katz2019}. 
Covariant PRT calculations  
  \citep[][]{Chan2019} 
  are therefore essential,  
  if we wish to take full account of 
  all the magneto-ionic plasma effects 
  throughout the evolutionary history of the Universe, 
  providing insights and
  theoretical bases 
  for proper interpretation of the statistical RM analyses 
  of the observed polarised sky.

\section{Results and Discussion: Variance of RM fluctuations}%
\label{sec-RM_var}

\subsection{Assessing the rotation measure fluctuation approach}
\label{subsec-RMF-formula}

The formula in equation~(\ref{eq-sigrm_choryu}) is commonly used 
 in RM fluctuation analysis 
 for probing the structures in 
 large-scale magnetic fields. 
Here we assess when the formula is justified
   and when it deserves caution.
The formula contains two variables related to $n_{\rm e}$ and $\boldsymbol{B}$,
  and our assessment will focus on their spatial distribution properties. 
We perform Monte-Carlo simulations 
  to compute the RM fluctuations. 
We consider simulated cubes of Mpc size 
  with mock thermal electron number density (d) 
  and magnetic field strength (b) with 
  uniform (U), Gaussian (G), fractal (F) and log-normal (L) distributions.
Each simulation is specified by a 4-letter label. 
For instance, `UdGb' stands for uniformly-distributed densities and gaussian-distributed magnetic field strengths with random orientations.  
The Mersenne Twister 
    \citep[MT,][]{Matsumoto1998} is implemented to generate uniformly-distributed pseudo-random numbers, 
    $Z \in (\;\!0, 1\;\!]$,
    which transform into the G, F and L distributed variates according to the specification.

The cubes are discretised into $256^3$ voxels, 
  each having an equal linear length $\overline{\Delta s}$ on the three sides. 
The magnetic field and the thermal electrons are specified according to the assigned distributions.  
Their values are normalised such 
  that they are of a similar order to those observed in galaxy clusters:
 $\langle n_{\rm e,th} \rangle = 10^{-3}\, {\rm cm}^{-3}$, 
    $B_{\rm rms} = 1 \,\mu {\rm G}$ and
        $L = 1 \,{\rm Mpc}$
        \citep[e.g.][]{Cho2009}. 
The total thermal electron number density and the total magnetic energy in the whole simulation box, 
   regardless of the magneto-ionic distribution, are
     $256^3 \times 10^{-3} = \, 16777.216 \, {\rm cm}^{-3} $ and 
     $|\boldsymbol{B}|^2 \,= 256^3 = 16777216 \, (\mu {\rm G})^2$, respectively.     
This ensures uniformity between the model cubes, 
  which enables
  direct comparisons between the simulations.

To compute the RM,  
  we sum the contributions along the lines-of-sight,
    $\boldsymbol{x}, \boldsymbol{y}$ and $\boldsymbol{z}$,    
    using the discretised expression of equation~(\ref{eq:RM_int})
in terms of lattice units $(i, j, k)$,
\begin{equation}
 {\cal R}_\perp  =  0.812
  \sum_\parallel  
   \frac{{\overline{\Delta s}}}{{\rm pc}} \!
     \left[ \left(\frac{n_{\rm e, th}(i,j,k)}{{\rm cm}^{-3}} \right)\! \left( \frac{B(i,j,k)}{\mu{\rm G}}\right)\right]_\parallel 
     {\rm rad}~{\rm m}^{-2} \  .
     \label{eq-RM_perp}
\end{equation}
The standard deviation $\varsigma_{_{\mathcal{R}\perp}}$ 
  across the simulated sky-plane 
  is then computed 
  and compared to the longitudinal standard deviation given in equation~(\ref{eq-sigrm_choryu}).

\subsubsection{Modelling magnetic fields}

We consider magnetic fields with random orientations and no spatial correlation.  
They are therefore unit vectors:    
    ${\hat B}_{\rm x} = \sin \theta \cos \phi$, 
    ${\hat B}_{\rm y} = \sin \theta \sin \phi$ and 
    ${\hat B}_{\rm z} = \cos \theta$,  
    with 
    $\cos\theta \in (\;\!-1, 1\;\!]$ and 
    $\phi \in (\;\!0, 2\pi\;\!]$ in a uniform distribution. 
The field strength on the other hand 
 has a uniform,
 non-solenoidal (Ub*)
 or a uniform, solenoidal (Ub)
 or a Gaussian (Gb) distribution.
The normalisation is such that the r.m.s. value 
 $B_{\rm rms} = 1 \,\mu {\rm G}$.  
Hence, we have 
  $B_{\parallel \rm{rms}} = B_{\rm{rms}}/\sqrt{3} \simeq 0.577 ~\mu \rm{G}$.
The Gaussian distribution is generated 
  using the Box-Muller transform 
  in the usual Monte-Carlo simulations. 
The simulated magnetic fields are then 
  cleaned in Fourier space 
  with the application of 
   a divergence-free ($\nabla \cdot \boldsymbol{B} = 0$) filter:
    $B_{i}(k_{m}) = 
    (\delta_{ij} - k_{i}k_{j}/k^2) \,
    {\tilde B}_{j}(k_{m})$
\citep{Balsara1998}.
As the process removes the field component parallel 
  to $\boldsymbol{k}$, 
   the total magnetic energy stored in the cube 
   shrinks to $2/3$ of its original value.  
To compensate for the energy loss by the filtering process, the field components are rescaled by a factor of $\sqrt{3/2}$.
An inverse Fourier transform is then conducted 
  to obtain the divergence-free (`solenoidal') magnetic field 
  in the configuration space  
  (see Appendix~\ref{app-div_free}).

We also note that the divergence-free filtering process 
  introduces a residual dipole, which has a preferred orientation,  
   depending on how the filtering process is executed.  
To suppress this dipole structure,\label{dipole}
  we employ a quick-fix solution\footnote{In a more proper treatment, 
   we would need 
   a superposition of three anti-parallel pairs of independent, orthogonal 
   field realisations in order to completely 
   remove the dipole. 
The process would then leave a residual quadrupole.  
Nonetheless the quick-fix solution that 
  we employ to suppress the dipole 
  is sufficient for the purpose of this demonstrative study.
In reality, the divergence filtering 
   is not always necessary before radiative transfer,  
   as the magnetic fields output by 
   a detailed magneto-hydrodynamic simulation 
   \citep[see e.g.][]{Marinacci2015,Marinacci2018,Barnes2018MNRAS}
   should be divergence free, at least in principle.}
using a superposition of three independent, orthogonal field realisations.  
We then renormalise the resultant  field from the superposition
  by a $\sqrt{1/3}$ scaling factor.
Since the realisations prepared as such 
  are divergence free in real (configuration) and Fourier space,
  their linear superpositions in real and Fourier space\footnote{Note that in the execution of Fourier transform process, 
   we do not consider an infinite span of the configuration space. 
The restriction of the electron number density and magnetic field structure within a finite volume 
  is equivalent to the introduction of a cubic window function 
  to an infinite configuration space. 
Thus, the density distribution and the magnetic field structure 
  that we obtain in the Fourier representation 
  are the convolutions of the cubic window function 
  with  the electron number density distribution and the magnetic field structure. 
  } 
  will also be divergence free.

\subsubsection{Modelling free-electron number density}

We consider four model electron number density distributions: 
  uniform (Ud), Gaussian (Gd), log-normal (Ld) and fractal (Fd). 
For the uniform distribution, the electron number density is set to be 1 unit in each cell.  
For the Gaussian distribution,
    we apply a Box-Muller transform,  
    setting the standard deviation to 0.2 times the mean 
    $10^{-3}\,{\rm cm}^{-3}$, 
    so that there are only a few negative numbers, which 
    can be converted to positives by simply taking the absolute. 
The log-normal distribution 
  is generated by taking the exponential function of Gaussian-distributed 
  random numbers.
For a fractal model, 
   we generate random phases $(k_{\rm x}, k_{\rm y}, k_{\rm z})$ in Fourier space. 
We then apply a power-law filter 
    $|\boldsymbol{k}|^{-5/3}$ 
	to mimic a 
	Kolmogorov-like turbulence spectrum
	(\citeyear{Kolmogorov1941a,Kolmogorov1941b}).
Simulations predict various kinds of turbulence
    in clusters and cosmic filaments
    \citep[e.g.][]{Iapichino2011}.
Scaling laws originally derived for incompressible media
    also turn out to be 
    a good approximation for compressible turbulence
    in subsonic regions of
    real observed or numerically simulated IGM
    \citep[see e.g.][]{Schuecker2004,Miniati2014,Nakwacki2016,White2019}.%
\footnote{%
The alternative extreme, of shock-compressed supersonic turbulence,
    yields steeper spectra $\sim k^{-2.1}$;
    e.g. \cite{Lee1991,Federrath2013}.}
In our model, we apply frequency cutoffs
    as in \cite{Saxton2005a}:
    we impose $k_\mathrm{max}=N/2$
    to prevent excessively sharp contrasts at voxel scale,
    and $k_\mathrm{min}=8$
    to prevent any single density peak dominating.
The inverse Fast Fourier Transform yields
   a fractal-like spatial structure
   with normally distributed local values $\mathcal N$.
Dimensionless positive densities are obtained by taking
    $\exp\;\!(\alpha\;\!\! {\cal N} / {\cal N}_{\rm max} )$, 
    where the contrast factor $\alpha = 4$
    and ${\cal N}_{\rm max}$ is a fiducial maximum fluctuation
        \citep{Elmegreen1989, Elmegreen2002}.
Lastly we obtain the various astrophysical configurations
   of thermal electron number densities
   by normalising $\langle n_{\rm e,th} \rangle$
   of each box to $10^{-3}\, {\rm cm}^{-3}$.


\subsubsection{RM dependence on the density and magnetic field structures}     

\begin{figure*}
\begin{minipage}{170mm}
    \centering
\includegraphics[width=0.7\textwidth]{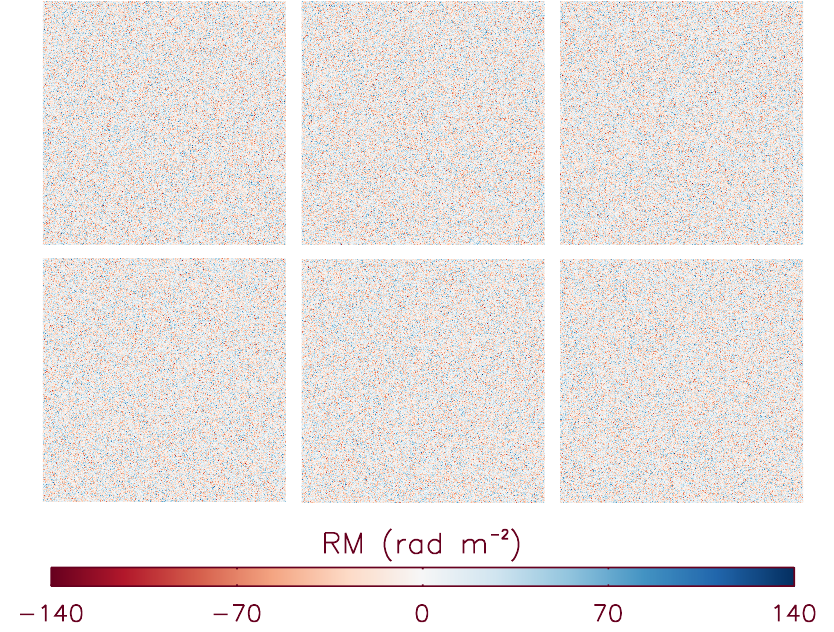}
    \caption{Synthetic RM maps of one realisation of GdGb (top) and UdUb* (bottom) in the xy (left), xz (middle) and yz (right) planes.}
    \label{fig:rm-map}
    \end{minipage}
\end{figure*}

We calculate synthetic RM maps by integrating along lines of sight $\boldsymbol{x}, \boldsymbol{y}$ and $\boldsymbol{z}$ using equation~(\ref{eq-RM_perp}) through various distributions of thermal electron number densities and magnetic field strengths.
The RM maps from the GdGb and UdUb* distributions are indistinguishable from a simple eyeball test (see Fig.~\ref{fig:rm-map}),
    even though the maps are generated from distinct distributions of number densities and magnetic field strengths.
GdGb is commonly assumed in astrophysical scenarios, whereas UdUb* is simply unrealistic because the magnetic fields are non-solenoidal.
The resulting RM maps are similar across all lines of sight, 
    implying that it is non trivial to characterise the thermal number densities and the magnetic field strengths from the observed RM fluctuations alone.    

We compare models quantitatively in
    Table~\ref{tab:rm-cluster}.
    We calculate 
    the line-of-sight longitudinal dispersions
    using equation~(\ref{eq-sigrm_choryu}) 
    and obtain
    $\sigma_{\cal R}^{\rm xy} \simeq \sigma_{\cal R}^{\rm xz} \simeq \sigma_{\cal R}^{\rm yz} \simeq 29.3 \, \rm{rad\,m^{-2}}$ 
    in all cases, 
    indicating
    that this type of RM fluctuation formula cannot distinguish between the different distributions of number densities and magnetic field strengths.
The tiny variations in the least significant figures of
    $\sigma_{\cal R}^{\rm xy}$, 
    $\sigma_{\cal R}^{\rm xz}$ and 
    $\sigma_{\cal R}^{\rm yz}$ 
are due to the numerical noise and random differences in the generated realisations.
Table~\ref{tab:rm-cluster} also shows that 
     line-of-sight and sky transverse fluctuations match reasonably well
     ($\sigma_{_\mathcal{R}}\simeq\varsigma_{_\mathcal{R}}$)
in the cases of UdUb*, UdUb and UdGb, 
    indicating that the widely-used RM fluctuation formula is applicable for uniformly-distributed densities and magnetic field strengths with uniform distributions and Gaussian distributions. 
However, for GdUb, GdGb, FdUb, FdGb, LdUb and LdGb; 
    $\sigma_{_\mathcal{R}}<\varsigma_{_\mathcal{R}}$,
    meaning that the RM fluctuation formula is inadequate in situations 
    with Gaussian, fractal or log-normal density distributions.
The disagreement between $\sigma_{_\mathcal{R}}$ and $\varsigma_{_\mathcal{R}}$
   is at the level of $\sim2\%$, $\sim25\%$ and $\sim40\%$
   respectively for G, F and L density models.
For a comparison,
   we note that
   \cite{Bhat2013}
   calculated the evolving RM properties
   of the ICM in fluctuation dynamo simulations,
   and found that $\varsigma_{_\mathcal{R}}$
   was $\sim10\%$--$15\%$ above some statistical indicators of RMF
   ($\approx\sigma_{_\mathcal{R}}$).
In their models, the evolving magnetic features
   seemed to be more influential than the density variations.

Notably, both our RMS model and explicit RT simulation 
    cannot distinguish the difference between
    solenoidal and non-solenoidal fields,
    as shown by 
    $\sigma_{_\mathcal{R}} ({\rm UdUb}^*) \simeq 
    \sigma_{_\mathcal{R}} ({\rm UdUb})
    \simeq 29.3 \, \rm{rad\,m^{-2}}$. 
Our calculations also show that 
    the sky planar fluctuations:
    $\varsigma_{_\mathcal{R}} ({\rm UdUb}^*) 
    \simeq \varsigma_{_\mathcal{R}} ({\rm UdUb}) 
    \simeq \varsigma_{_\mathcal{R}} ({\rm UdGb})$,
    $\varsigma_{_\mathcal{R}} ({\rm GdUb}) 
    \simeq \varsigma_{_\mathcal{R}} ({\rm GdGb})$,
    $\varsigma_{_\mathcal{R}} ({\rm FdUb}) 
    \simeq \varsigma_{_\mathcal{R}} ({\rm FdGb})$
    and
    $\varsigma_{_\mathcal{R}} ({\rm LdUb}) 
    \simeq \varsigma_{_\mathcal{R}} ({\rm LdGb})$.
 The Fourier transform and inverse Fourier transform
    are part of the divergence cleaning process. 
Note that the Fourier transform of uniformly-distributed fields in a finite volume 
  gives a 3-D sinc function (in the Cartesian coordinate).

\begin{figure*}
\begin{minipage}{170mm}  
    \centering
\includegraphics[width=0.64\textwidth]{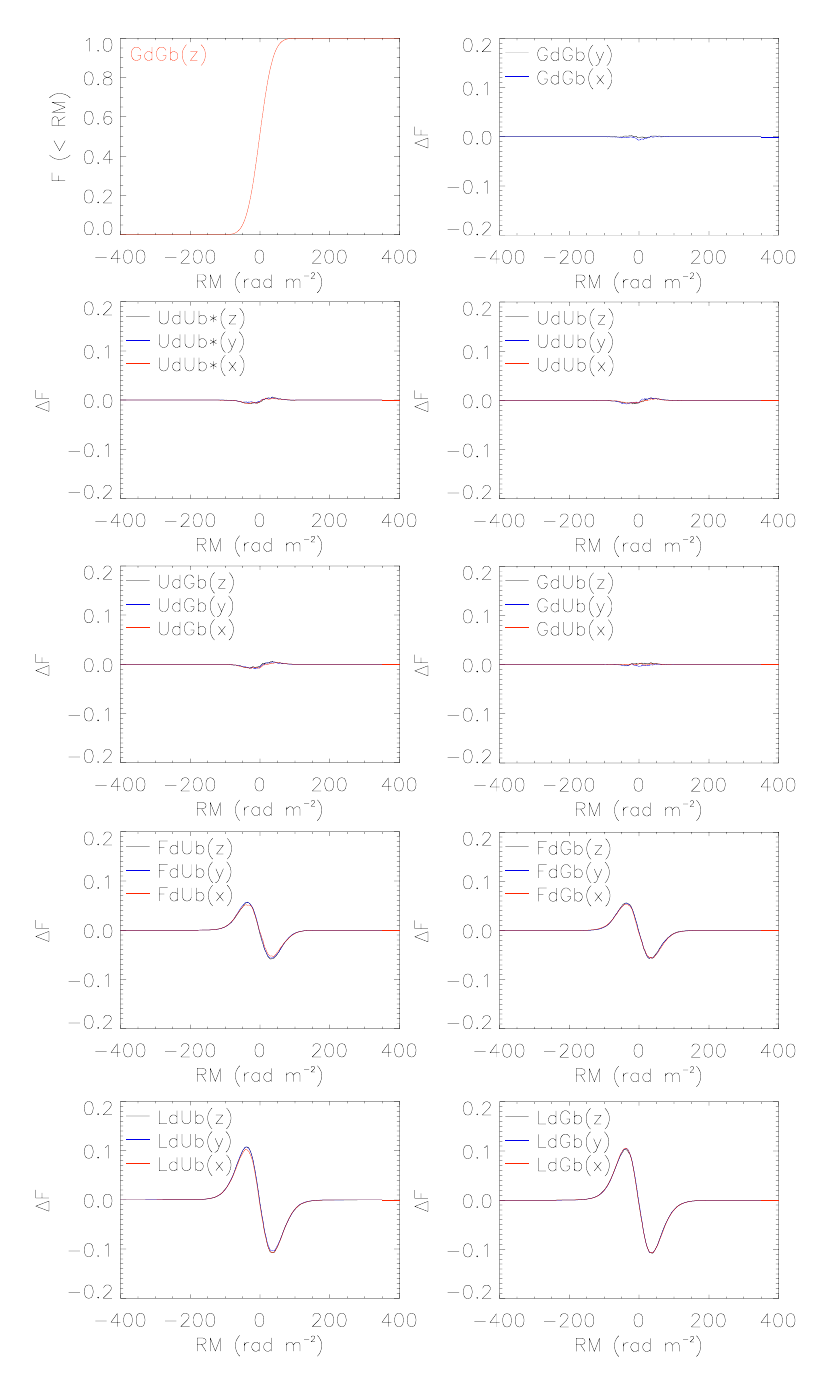}
    \caption{Top-left panel showing the CDF of GdGb(z) as a reference to calculate the numerical differences with the CDFs of GdGb(y), GdGb(x) as well as UdUb*, UdUb, UdGb, GdUb, FdUb, FdGb, LdUb and LdGb for every line-of-sight. The CDFs are almost identical in all cases, except for models with fractal and log-normal density distributions.}
      \label{fig:rm-cdf}
    \end{minipage}
\end{figure*}

For a more detailed characterisation of the RM distributions,
    we calculate the histograms and 
    cumulative distribution functions (CDFs) 
    of the RM maps from every line of sight 
    through the UdUb*, UdUb, UdGb, GdUb, GdGb, FdUb, FdGb, LdUb and LdGb distributions.
For each cube,
    the histograms and CDFs along lines-of-sight
    $\boldsymbol{x}, \boldsymbol{y}$ and $\boldsymbol{z}$
    coincide, 
    confirming that isotropy is preserved in each box
    (also demonstrated by the results in Table~\ref{tab:rm-cluster}, where  
    $\sigma_{\cal R}^{\rm xy} \simeq \sigma_{\cal R}^{\rm xz} \simeq \sigma_{\cal R}^{\rm yz}$ 
    for all simulations).
We set GdGb(z) as the basis CDF 
    and calculate its numerical difference from 
    the CDFs of the rest of the models. 
The CDFs at every line-of-sight are almost indistinguishable 
    in each case of GdGb, UdUb*, UdUb, UdGb and GdUb, 
    as shown by the tiny fluctuations in the zero line (Fig.~\ref{fig:rm-cdf}).
The CDFs of FdUb, FdGb, LdUb and LdGb deviate significantly from the basis CDF. 

Using the numerical CDF curves,
    we perform a Kolmogorov-Smirnov (KS) test with the null hypothesis being that the two 
    RM
    samples,
    observed either in the $\boldsymbol{x}$, $\boldsymbol{y}$ or $\boldsymbol{z}$ direction,
    are drawn from the same distribution.
The KS test is non-parametric and reports the maximum value of absolute (vertical) difference between two CDFs
    \citep[see e.g.][]{Press2007}.
We 
calculate the KS statistics 
which are summarised in Table~\ref{tab:ks-cluster}.
We obtain $D \ll 1$ and p-value probabilities in the range of $0.2 - 0.6$, favouring the null hypothesis since $p > 0.05$.
    Our KS tests do not show evidence of anisotropy.

\begin{figure*}
\begin{minipage}{170mm}
    \centering
\includegraphics[width=0.7\textwidth]{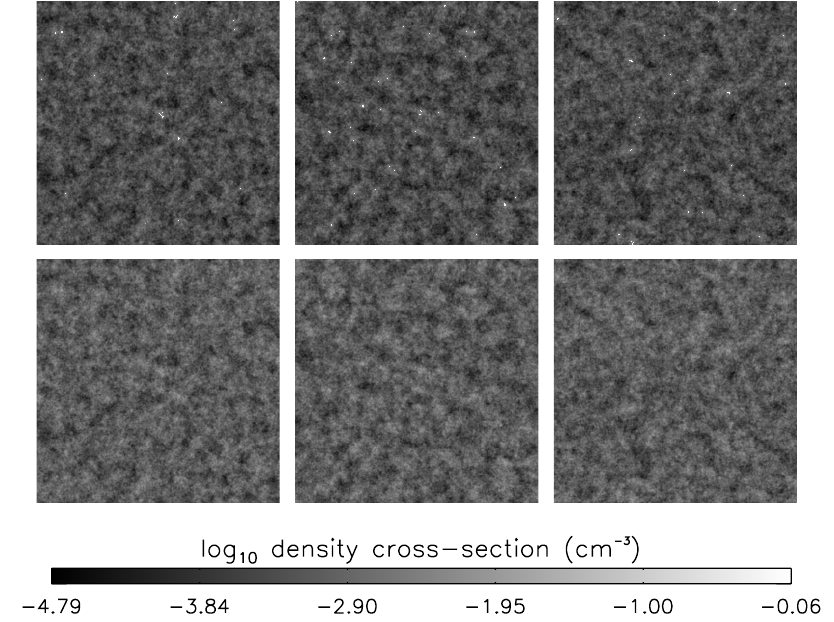}
    \caption{Cross-sections of Cd3 (top) and Cd5 (bottom) at x = 128 (left), y = 128 (middle) and z = 128 (right).}
      \label{fig:cloudy-cs}
    \end{minipage}
\end{figure*}

\begin{figure*}
\begin{minipage}{170mm}
    \centering
\includegraphics[width=0.7\textwidth]{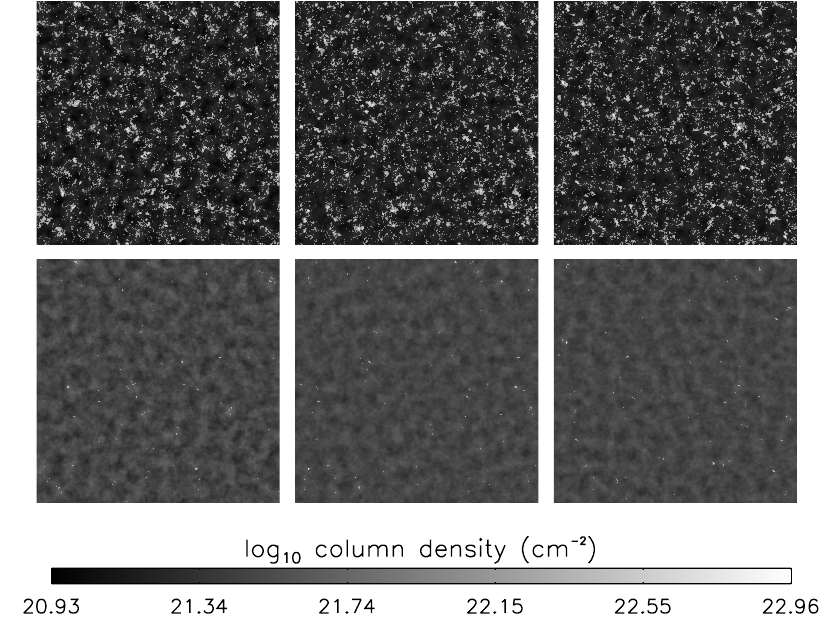}
    \caption{Column densities of Cd3 (top) and Cd5 (bottom) in the x (left), y (middle) and z (right) directions.}
      \label{fig:cloudy-proj}
    \end{minipage}
\end{figure*}

\begin{figure*}
\begin{minipage}{170mm}
    \centering
    \vspace{2cm}
\includegraphics[width=0.7\textwidth]{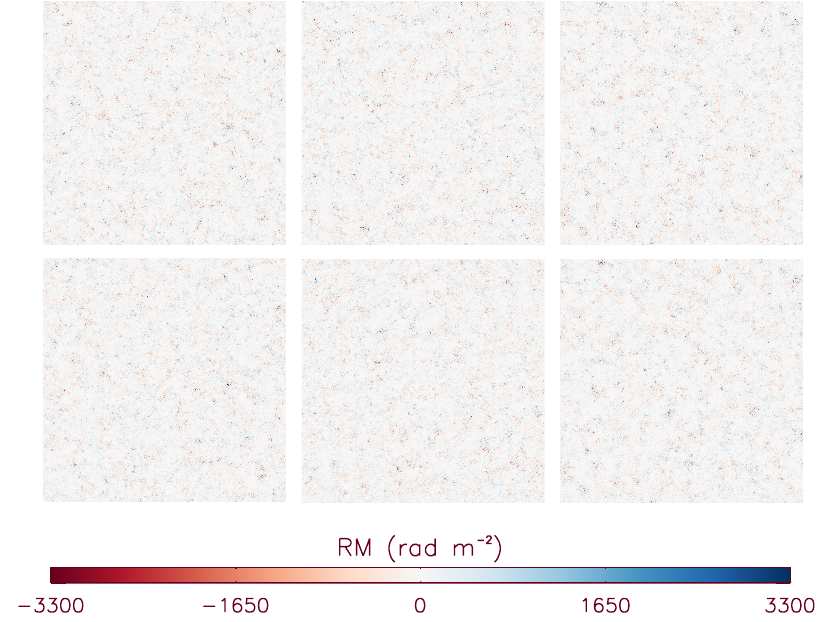}
    \caption{Synthetic RM maps of Cd2Gb (top) and Cd2Ub* (bottom) in the xy (left), xz (middle) and yz (right) planes. }
      \label{fig:rm-cloudy-map}
    \end{minipage}
\end{figure*}

In addition, we consider a fractal medium with two density phases 
   (hereafter referred to as cloudy models), 
   mimicking the typical environments in the ICM/ISM
    (see Appendix~\ref{app-cloudy_blocks}).
We consider various cloud volume filling factors 
$f = 10^{-2}$,     
    $10^{-3}$,
    $10^{-4}$,
    $10^{-5}$, and
    $10^{-6}$, 
    corresponding to 
Cd2, Cd3, Cd4, Cd5 and Cd6, respectively.
Figs.~\ref{fig:cloudy-cs} and \ref{fig:cloudy-proj} show the log$_{10}$ cross-sections and column densities of Cd3 and Cd5, respectively.  
The cross-sections are a slice taken from the cloudy models Cd3 and Cd5 at $x = 128$, $y = 128$ and $z = 128$.
The cross-sections and column densities show the non-uniformity of the diffuse media and the cloud phases. 
The cloudy models are fairly isotropic along every line-of-sight, with Cd3 being more dense than Cd5, as indicated by the larger number of bright specks embedded within the cloudy media.
    Fig.~\ref{fig:rm-cloudy-map} shows that the RM maps of the Cd2Gb and Cd2Ub* distributions are indistinguishable, despite the distinction between the distributions of magnetic field strengths, especially with Ub* being non-solenoidal and unphysical. 
    Moreover, our
calculations in Table~\ref{tab:rm-cloudy} show that the RMS statistics are unable to tell the cloudy features apart, in spite of Cd2 -- 6 having different volume filling factors. 
In particular, the RMS statistics for various distributions of Cd and b are 
    $\sigma_{_\mathcal{R}} \simeq 29.3$\, rad m$^{-2}$,
    which is similar to the RMS statistics for various distributions of our single-phase models in Table~\ref{tab:rm-cluster},
    indicating that the RMS method cannot distinguish between 
    a range of different clumpy (or smooth) configurations of density
    and magnetic fields.
We also calculate the sky-transverse standard deviations and find that, 
    with the exception of the overcast model Cd2, the 
    $\varsigma_{_\mathcal{R}}$ 
    decreases with decreasing volume filling factor. 
This is expected since the scatter should be less with fewer clouds
    (and we approach the L lognormal models as $f\rightarrow0$).
Clumpiness always causes $\varsigma_{_\mathcal{R}}>\sigma_{_\mathcal{R}}$,
   and often by large multiples
   (with relative differences up to $94\%$).
Furthermore, 
    comparing the dispersions between the longitudinal direction and the sky transverse direction,
    the cloudy models in Table~\ref{tab:rm-cloudy} show a greater scatter 
    than the Ud, Gd, Fd and Ld models did in Table~\ref{tab:rm-cluster}. 
The variability of standard deviations 
    may be attributed to the random shapes and orientations of the clouds. 

We also calculate histograms (not shown)
    to characterise the RM distributions
    from every line-of-sight 
    through the
    Cd2Ub*, Cd2Ub, Cd2Gb, Cd3Ub*, Cd3Ub, Cd3Gb, Cd4Ub*, Cd4Ub, Cd4Gb, Cd5Ub*, Cd5Ub, Cd5Gb, Cd6Ub*, Cd6Ub and Cd6Gb
    distributions.
For each cube,
    the histograms along lines-of-sight
    $\boldsymbol{x}, \boldsymbol{y}$ and $\boldsymbol{z}$
    coincide, 
    confirming that isotropy is preserved in each box, 
    which is also shown by 
    $\varsigma_{\cal R}^{\rm xy} \simeq \varsigma_{\cal R}^{\rm xz} \simeq \varsigma_{\cal R}^{\rm yz}$ 
    in Table~\ref{tab:rm-cloudy}. 

\begin{figure*}
\begin{minipage}{170mm}
    \centering
\includegraphics[width=0.95\textwidth]{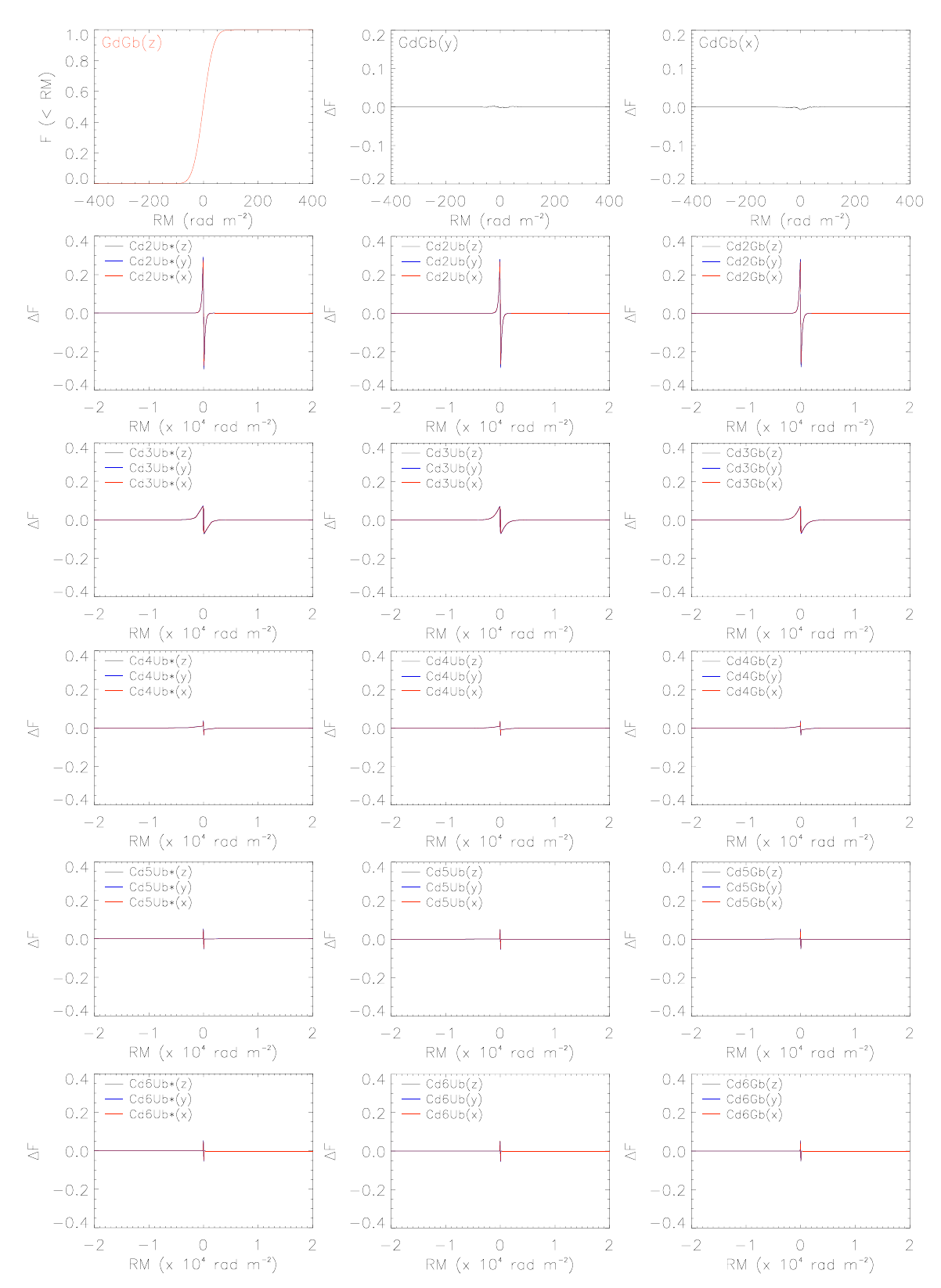}
    \caption{Top-left panel showing the CDF of GdGb(z) as a reference to calculate the numerical differences with the CDFs of GdGb(y), GdGb(x) as well as Cd2Ub*, Cd2Ub, Cd2Gb, Cd3Ub*, Cd3Ub, Cd3Gb, Cd4Ub*, Cd4Ub, Cd4Gb, Cd5Ub*, Cd5Ub, Cd5Gb, Cd6Ub*, Cd6Ub and Cd6Gb at every line-of-sight.}
      \label{fig:rm-cloudy-cdf-1}
    \end{minipage}
\end{figure*}

    The results from the KS-tests are summarised in Table~\ref{tab:ks-cloudy}. The KS statistics do not show evidence of anisotropy.
Using GdGb(z) as the basis CDF,
    we calculate its numerical difference from 
    the CDFs of the cloudy, magnetised models
    in 
    Fig.~\ref{fig:rm-cloudy-cdf-1}.
These panels are almost indistinguishable between 
    different configurations of magnetic fields, for example, the numerical difference trends for Cd3Ub*(x,y,z), 
    Cd3Ub(x,y,z) and 
    Cd3Gb(x,y,z) look similar to each other.
    This suggests that the RMs are more dependent on the cloudy structures, rather 
    than the magnetic field configurations
    since the density variations are of orders of magnitude within each cube,
    while the dynamic variation of the magnetic field is relatively smaller. 
Notably the CDFs of Cd2 show the largest deviation from the basis CDF, whereas the CDFs for the rest of the models, apart from Cd3, are almost indistinguishable. 
This may be a consequence of a scarcity of clouds in Cd4, 5 and 6. 
    
Hence, from our results above, the widely-used RM fluctuation formula (RMS statistics) is valid when all of
the following conditions hold:
(i) a random field produces random Faraday rotation,
(ii) there exists a meaningful characteristic thermal electron number density,
(iii) there exists a uniform or Gaussian distribution of magnetic field strengths,
(iv) the field is isotropic, and
(v) the density and the magnetic field are not correlated.
In situations where 
some of these criteria are not met, the RMS statistics would be inadequate to be used to interpret the magnetic field properties from the RM analyses.
Discrepancies could in principle be large in some environments such as
    cluster cores where multiphase features are obvious in other wavebands
    \citep[e.g.][]{Conselice2001}
    or faint cluster outskirts where clumpiness is conjectured
    \citep[e.g.][]{Urban2014}.


\begin{table*}
    \centering
    \begin{tabular}{clr@{.}lr@{.}lr@{.}lr@{.}lr@{.}lr@{.}lr@{.}lr@{.}lr@{.}lccc}
\hline
\\
\multicolumn{2}{c}{distribution}
&\multicolumn{6}{c}{longitudinal, eq.~(\ref{eq-sigrm_choryu})}
&\multicolumn{6}{c}{sky transverse, eq.~(\ref{eq-RM_perp})}
&\multicolumn{6}{c}{ratio}
\\
\\
&
&\multicolumn{2}{c}{$\sigma_{\mathcal{R}}^\mathrm{xy}$}
&\multicolumn{2}{c}{$\sigma_{\mathcal{R}}^\mathrm{xz}$}
&\multicolumn{2}{c}{$\sigma_{\mathcal{R}}^\mathrm{yz}$}
&\multicolumn{2}{c}{$\varsigma_{\mathcal{R}}^\mathrm{xy}$}
&\multicolumn{2}{c}{$\varsigma_{\mathcal{R}}^\mathrm{xz}$}
&\multicolumn{2}{c}{$\varsigma_{\mathcal{R}}^\mathrm{yz}$}
&\multicolumn{2}{c}{$\sigma^\mathrm{xy}/\varsigma^\mathrm{xy}$}
&\multicolumn{2}{c}{$\sigma^\mathrm{xz}/\varsigma^\mathrm{xz}$}
&\multicolumn{2}{c}{$\sigma^\mathrm{yz}/\varsigma^\mathrm{yz}$}
\\
\\
\hline
\\
Ud & Ub*
& 29&2971 & 29&2975 & 29&2962
& 29&2861 & 29&3151 & 29&3042
& 1&0004 & 0&9994 & 0&9997
\\
Ud & Ub
& 29&2962 & 29&2977 & 29&2970
& 29&3101 & 29&3198 & 29&3278
& 0&9995 & 0&9993 & 0&9990
\\
Ud & Gb
& 29&2965 & 29&2965 & 29&2979
& 29&3231 & 29&2934 & 29&3299
& 0&9991 & 1&0001 & 0&9989
\\
\\
Gd & Ub
& 29&2982 & 29&2961 & 29&2966
& 29&9263 & 29&8972 & 29&8994
& 0&9790 & 0&9799 & 0&9798
\\
Gd & Gb
& 29&2970 & 29&2973 & 29&2966
& 29&9113 & 29&8854 & 29&8902
& 0&9795 & 0&9803& 0&9801
\\
\\
Fd & Ub
& 29&2987 & 29&2965 & 29&2956
& 39&1187 & 39&1392 & 39&1134
& 0&7490 & 0&7485 & 0&7490
\\
Fd & Gb
& 29&2975 & 29&2966 & 29&2968
& 39&1185 & 39&1357 & 39&1186
& 0&7489 & 0&7486 & 0&7489
\\
\\
Ld & Ub
& 29&2969 & 29&2968 & 29&2972
& 48&3524 & 48&3218 & 48&3327
& 0&6059 & 0&6063 & 0&6062
\\
Ld & Gb
& 29&2975 & 29&2964 & 29&2970
& 48&3058 & 48&3017 & 48&3146
& 0&6065 & 0&6065 & 0&6064
\\
\\
\hline
    \end{tabular}
    \caption{
    The dispersion of RM,
    $\sigma_{\mathcal{R}}^{\rm xy}$,
    $\sigma_{\mathcal{R}}^{\rm xz}$ and
    $\sigma_{\mathcal{R}}^{\rm yz}$,
    calculated along lines-of-sight z, y, and x, respectively,
    as light travels through various configurations
    of magnetised, thermal plasma.
    The transverse dispersion 
    is calculated from the RM maps over the sky plane:
    $\varsigma_{\mathcal{R}}^{\rm xy}$,
    $\varsigma_{\mathcal{R}}^{\rm xz}$ and
    $\varsigma_{\mathcal{R}}^{\rm yz}$.
    The last three columns show
    the ratios of the longitudinal dispersion to the transverse dispersion. 
    The realisations are normalised 
    to the order of magnitude typically found in galaxy clusters:
    $\langle n_{\rm e,th} \rangle \sim 10^{-3}\, {\rm cm}^{-3}$, $B_{\rm rms} \sim 1 \,\mu {\rm G}$ and
    $L \sim 1 \,{\rm Mpc}$.
    The magnetic fields are strictly divergence-free, except for Ub*.}
    \label{tab:rm-cluster}
\end{table*}

\begin{table*}
    \centering
    \begin{tabular}{clr@{.}lr@{.}lr@{.}lr@{.}lr@{.}lr@{.}lr@{.}lr@{.}lr@{.}lccc}
\hline
\\
\multicolumn{2}{c}{distribution}
&\multicolumn{6}{c}{$D$}
&\multicolumn{6}{c}{$p$-value}
\\
\\
&
&\multicolumn{2}{c}{xy \& xz}
&\multicolumn{2}{c}{xy \& yz}
&\multicolumn{2}{c}{xz \& yz}
&\multicolumn{2}{c}{xy \& xz}
&\multicolumn{2}{c}{xy \& yz}
&\multicolumn{2}{c}{xz \& yz}
\\
\\
\hline
\\
Ud & Ub*
& 0&0044 & 0&0048 & 0&0045  
& 0&5707 & 0&4763 & 0&5460
\\
Ud & Ub
& 0&0052 & 0&0047 & 0&0052
& 0&4272 & 0&5502 & 0&4364
\\
Ud & Gb
& 0&0052 & 0&0055 & 0&0057
& 0&4255 & 0&3654 & 0&3712 
\\
\\
Gd & Ub
& 0&0052 & 0&0048 & 0&0052 
& 0&4240 & 0&5123 & 0&4509 
\\
Gd & Gb
& 0&0050 & 0&0051 & 0&0051 
& 0&4696 & 0&4359 & 0&4275
\\
\\
Fd & Ub
& 0&0047 & 0&0072 & 0&0072  
& 0&5110 & 0&1549 & 0&1421 
\\
Fd & Gb
& 0&0049 & 0&0068 & 0&0071 
& 0&5052 & 0&1696 & 0&1658 
\\
\\
Ld & Ub
& 0&0050 & 0&0050 & 0&0050 
& 0&4833 & 0&4662 & 0&4644 
\\
Ld & Gb
& 0&0049 & 0&0045 & 0&0049
& 0&4806 & 0&5550 & 0&4865 
\\
\\
\hline
    \end{tabular}
    \caption{The KS statistics $D$ and $p$-value probabilities corresponding to various configurations in Table~\ref{tab:rm-cluster}.
    }
    \label{tab:ks-cluster}
\end{table*}

\begin{table*}
    \centering
    \begin{tabular}{clr@{.}lr@{.}lr@{.}lr@{.}lr@{.}lr@{.}lr@{.}lr@{.}lr@{.}lccc}
\hline
\\
\multicolumn{2}{c}{distribution}
&\multicolumn{6}{c}{longitudinal, eq.~(\ref{eq-sigrm_choryu})}
&\multicolumn{6}{c}{sky transverse, eq.~(\ref{eq-RM_perp})}
&\multicolumn{6}{c}{ratio}
\\
\\
&
&\multicolumn{2}{c}{$\sigma_{\mathcal{R}}^\mathrm{xy}$}
&\multicolumn{2}{c}{$\sigma_{\mathcal{R}}^\mathrm{xz}$}
&\multicolumn{2}{c}{$\sigma_{\mathcal{R}}^\mathrm{yz}$}
&\multicolumn{2}{c}{$\varsigma_{\mathcal{R}}^\mathrm{xy}$}
&\multicolumn{2}{c}{$\varsigma_{\mathcal{R}}^\mathrm{xz}$}
&\multicolumn{2}{c}{$\varsigma_{\mathcal{R}}^\mathrm{yz}$}
&\multicolumn{2}{c}{$\sigma^\mathrm{xy}/\varsigma^\mathrm{xy}$}
&\multicolumn{2}{c}{$\sigma^\mathrm{xz}/\varsigma^\mathrm{xz}$}
&\multicolumn{2}{c}{$\sigma^\mathrm{yz}/\varsigma^\mathrm{yz}$}
\\
\\
\hline
\\
Cd2 & Ub*
& 29&2976 & 29&2984 & 29&2949
& 280&6983 & 280&0177 & 280&4870
& 0&1044 & 0&1046 & 0&1044
\\
Cd2 & Ub
& 29&2922 & 29&2979 & 29&3008
& 280&5404 & 279&4889 & 281&2761
& 0&1044 & 0&1048 & 0&1042
\\
Cd2 & Gb
& 29&2968 & 29&2986 & 29&2956
& 279&8605 & 281&7843 & 280&9985
& 0&1047 & 0&1040 & 0&1043
\\
\\
Cd3 & Ub*
& 29&2976 & 29&2984 & 29&2949
& 471&8747 & 483&3400 & 473&8896
& 0&0621 & 0&0606 & 0&0618
\\
Cd3 & Ub
& 29&2922 & 29&2979 & 29&3008
& 478&3895 & 479&4463 & 476&4639
& 0&0612 & 0&0611 & 0&0615
\\
Cd3 & Gb
& 29&2968 & 29&2986 & 29&2956
& 466&7194 & 480&4994 & 475&2179
& 0&0628 & 0&0610 & 0&0617
\\
\\
Cd4 & Ub*
& 29&2976 & 29&2984 & 29&2949
& 267&1094 & 271&5542 & 265&8259
& 0&1100 & 0&1079 & 0&1102
\\
Cd4 & Ub
& 29&2922 & 29&2979 & 29&3008
& 276&8559 & 266&4851 & 267&9559
& 0&1058 & 0&1099 & 0&1093
\\
Cd4 & Gb
& 29&2968 & 29&2986 & 29&2956
& 257&7923 & 271&5684 & 268&0812
& 0&1136 & 0&1079 & 0&1093
\\
\\
Cd5 & Ub*
& 29&2976 & 29&2984 & 29&2949
& 93&4944 & 86&3299 & 89&4997
& 0&3134 & 0&3394 & 0&3273
\\
Cd5 & Ub
& 29&2922 & 29&2979 & 29&3008
& 98&3658 & 88&2455 & 97&2471
& 0&2978 & 0&3320 & 0&3013
\\
Cd5 & Gb
& 29&2968 & 29&2986 & 29&2956
& 83&6243 & 83&0571 & 89&1919
& 0&3503 & 0&3528 & 0&3285
\\
\\
Cd6 & Ub*
& 29&2976 & 29&2984 & 29&2949
& 44&5203 & 43&3651 & 43&0838
& 0&6581 & 0&6756 & 0&6800
\\
Cd6 & Ub
& 29&2922 & 29&2979 & 29&3008
& 46&9690 & 43&8330 & 45&7565
& 0&6236 & 0&6684 & 0&6404
\\
Cd6 & Gb
& 29&2968 & 29&2986 & 29&2956
& 42&4488 & 40&6938 & 44&7132
& 0&6902 & 0&7200 & 0&6552
\\
\\
\hline
    \end{tabular}
    \caption{The standard deviations of RM,
    $\sigma_{\mathcal{R}}^{\rm xy}$,
    $\sigma_{\mathcal{R}}^{\rm xz}$ and
    $\sigma_{\mathcal{R}}^{\rm yz}$,
    calculated
    along lines-of-sight z, y, and x, respectively, as light travels
    through various configurations of magnetised, two-phase fractal plasma.
The realisations are normalised to
    the order of magnitude typically found in galaxy clusters:
    $\langle n_{\rm e,th} \rangle \sim 10^{-3}\, {\rm cm}^{-3}$, $B_{\rm rms} \sim 1 \,\mu {\rm G}$ and
    $L \sim 1 \,{\rm Mpc}$.
The magnetic fields are strictly divergence-free, except for Ub*.
The volume filling factors for Cd2, Cd3, Cd4, Cd5 and Cd6 are
    $10^{-2}$,
    $10^{-3}$,
    $10^{-4}$,
    $10^{-5}$, and
    $10^{-6}$
    respectively.}
    \label{tab:rm-cloudy}
\end{table*}

\begin{table*}
    \centering
    \begin{minipage}{175mm}
    \centering
    \begin{tabular}{clr@{.}lr@{.}lr@{.}lr@{.}lr@{.}lr@{.}lr@{.}lr@{.}lr@{.}lccc}
\hline
\\
\multicolumn{2}{c}{distribution}
&\multicolumn{6}{c}{$D$}
&\multicolumn{6}{c}{$p$-value}
\\
\\
&
&\multicolumn{2}{c}{xy \& xz}
&\multicolumn{2}{c}{xy \& yz}
&\multicolumn{2}{c}{xz \& yz}
&\multicolumn{2}{c}{xy \& xz}
&\multicolumn{2}{c}{xy \& yz}
&\multicolumn{2}{c}{xz \& yz}
\\
\\
\hline
\\
Cd2 & Ub* 
& 5&7068E-03 & 1&9516E-02 & 1&9180E-02
& 2&3559E-01 & 2&7937E-11 & 6&5523E-11
\\
Cd2 & Ub 
& 6&3782E-03 & 2&0584E-02 & 2&3743E-02
& 1&3851E-01 & 1&6807E-12 & 1&7189E-16
\\
Cd2 & Gb 
& 4&2114E-03 & 1&7532E-02 & 2&0309E-02    
& 6&0557E-01 & 3&4726E-09 & 3&5123E-12
\\
\\
Cd3 & Ub* 
& 4&5929E-03 & 7&0648E-03 & 8&1940E-03 
& 4&9313E-01 & 7&5602E-02 & 2&4407E-02
\\
Cd3 & Ub 
& 6&5002E-03 & 5&8136E-03 & 6&1188E-03
& 1&2494E-01 & 2&1739E-01 & 1&7129E-01
\\
Cd3 & Gb 
& 3&6011E-03 & 8&2855E-03 & 7&5684E-03 
& 7&8844E-01 & 2&2106E-02 & 4&6618E-02
\\
\\
Cd4 & Ub* 
& 3&1281E-03 & 4&2267E-03 & 5&7831E-03
& 9&0517E-01 & 6&0094E-01 & 2&2247E-01
\\
Cd4 & Ub 
& 4&6844E-03 & 3&0975E-03 & 6&0120E-03
& 4&6754E-01 & 9&1125E-01 & 1&8646E-01
\\
Cd4 & Gb 
& 3&4027E-03 & 6&3477E-03 & 6&3019E-03
& 8&4190E-01 & 1&4208E-01 & 1&4758E-01
\\
\\
Cd5 & Ub* 
& 3&0823E-03 & 4&3945E-03 & 5&8289E-03
& 9&1421E-01 & 5&5062E-01 & 2&1488E-01
\\
Cd5 & Ub 
& 4&7150E-03 & 3&2654E-03 & 6&3171E-03
& 4&5916E-01 & 8&7535E-01 & 1&4573E-01
\\
Cd5 & Gb 
& 3&3112E-03 & 6&0883E-03 & 5&3864E-03
& 8&6457E-01 & 1&7552E-01 & 2&9698E-01
\\
\\
Cd6 & Ub* 
& 3&0060E-03 & 4&5624E-03 & 5&9357E-03
& 9&2823E-01 & 5&0180E-01 & 1&9792E-01
\\
Cd6 & Ub 
& 4&8370E-03 & 3&2043E-03 & 6&2866E-03
& 4&2644E-01 & 8&8908E-01 & 1&4945E-01
\\
Cd6 & Gb 
& 3&3112E-03 & 5&9357E-03 & 5&2948E-03
& 8&6457E-01 & 1&9792E-01 & 3&1644E-01
\\
\\
\hline
    \end{tabular}
    \caption{The KS statistics $D$ and $p$-value probabilities corresponding to the configurations in Table~\ref{tab:rm-cloudy}.}
    \label{tab:ks-cloudy}
\end{minipage}
\end{table*}

\subsection{Interpreting magnetic field properties from polarisation analyses}

\subsubsection{Ambiguity in the polarisation angle}
\label{subsec-n_pi}

The inference of RM from observations of linear polarisation is subjected 
 to an $n \uppi$ ambiguity in its direction
    \citep[][]{Ruzmaikin1979}.
For a clean line-of-sight 
  with a single point source, 
  the polarisation angle $\varphi$ 
  and the wavelengths $\lambda$ 
  satisfy a relationship: 
        $\varphi= \varphi_{0} + {\mathcal R}\lambda^{2}$, 
    fitting the observation for the intrinsic polarisation angle $\varphi_0$
    and the slope $\mathcal{R}$ 
    gives the rotation measure.
The foreground magnetic field structure 
    can be inferred from the RM
    if the emission measure is known.
In practice, the measured polarisation angle
    $\varphi$ can only be constrained between $0$ and $\uppi$, 
    hence there is an ambiguity of $\pm n \uppi$, where $n$ is an integer,  
    thus causing a problem 
    in determining 
    $\varphi_0$ and $\mathcal{R}$.

Early efforts were taken to
    resolve this ambiguity by imposing 
    a search limit for the best RM 
    from an astrophysical perspective
    and carrying out 
    observations in several frequencies
    so to obtain the best fit 
    using a chi-squared minimisation
        \citep[see e.g.][]{Simard1981, Rand1994}.
This method assumed that no $n \uppi$ ambiguity occurs between two closely-spaced wavelengths,
    such that $|\Delta \varphi| < \uppi / 2$ is fulfilled
    \citep{Ruzmaikin1979}.   
The source is observed across a radio broad band
   with sparsely sampled wavelengths, 
   and near each observed wavelength, 
   combinations of $(\varphi \pm n \uppi)$ 
   are considered in the fitting process.
While this method can be applied 
  to Faraday-thin media
  with a bright background point source, 
  it sometimes gives multiple acceptable solutions. 
It does not work well for faint sources. 
In the Faraday-thick regime, 
  the method will break down 
  because the linear relation above 
  does not hold. 
It is also problematic 
  when there are multiple sources along a line-of-sight 
   or when Faraday depolarisation occurs 
    \citep[see e.g.][]{Vallee1980,Sokoloff1998, Farnsworth2011}.

Recently, alternative methods have been 
  developed, for example,  
    the circular statistical method 
    \citep{Sarala2001},
    the \texttt{PACERMAN} algorithm
        \citep{Dolag2005, Vogt2005},
 the RM synthesis/\texttt{RMCLEAN} method
            \citep{Burn1966, Brentjens2005, Heald2009},
            Stokes $QU$-fitting
            \citep[e.g.][]{Farnsworth2011, OSullivan2012},
            and the dependence on RM of neighbouring sources 
            \citep{Taylor2009,Ma2019}.  
The $n \uppi$ ambiguity is one of the obstacles  
   that must be overcome 
   when analysing 
   large-scale magnetic fields using the RM information. 
On the other hand, we may bypass our reliance 
  on the RM statistics 
  by carrying out a proper (covariant) 
  polarised radiative transfer, 
  which can directly track the evolution 
  of polarisation along a line-of-sight
  to resolve the $n\uppi$ ambiguity
  without having to make an a priori assumption on the Faraday complexity 
  \citep[see][]{Chan2019}.

\subsubsection{Issues in analyses of 
  polarisation associated with large-scale astrophysical structures}
\label{subsec-scale}

\noindent
{\it FRBs and quasars as diagnostics:} \\ 
FRBs and quasars are exceptionally bright, 
    polarised radio sources, 
    observable 
    across cosmological distances.      
They are therefore useful probes 
  of the intergalactic magnetic fields 
    \citep[see e.g.][]{Xu2014redshift, Zheng2014, Akahori2016, Ravi2016, Vazza2018b, Hackstein2019}  and their evolution 
\citep[see e.g.][]{Xu2014redshift}, 
   if their redshifts 
    and dispersion measures are known
    \citep[see e.g.][]{KronbergPerry1982, Blasi1999, Kronberg2008, Xu2014b, Petroff2016}. 
Circular polarisation was detected in some quasars 
    \citep[see e.g.][]{Roberts1975, Saikia1988, Rayner2000, OSullivan2013} 
and FRBs
    \citep[see e.g.][]{Petroff2015, Petroff2017}, 
    indicating that Faraday conversion 
    \citep[see e.g.][]{Vedantham2019, Gruzinov2019}
    or scintillation-induced variations 
    \citep{Macquart2000}
    might occur.
As the number of detections of FRBs and quasars increases
    \citep[see e.g.][]{Keane2018Nature},
    the polarisation properties of their signals can be used to better constrain large-scale magnetic field properties.
Apart from the effects of Faraday conversion and scintillation,
    it is also important to 
    distinguish between the RM contributions from multiple sources along the line-of-sight,
    consider the effects of traversing multi-phase media, 
        as well as
    taking into account of the structural evolution and stretching of radiation wavelength in an expanding Universe 
    \citep[see e.g.][]{Han2017}.
In these situations, 
    RM is no longer sufficient to fully characterise the changes in polarisation
    and hence a covariant cosmological polarised radiative transfer treatment is necessary
    \citep[see][]{Chan2019}. 

\vspace*{0.3cm}  

\noindent
{\it Direct radio emission from large-scale structure:} \\ 
An emissive and Faraday-rotating medium 
    will result in a net depolarisation due to differential Faraday rotation \citep[e.g.][]{Sokoloff1998, Beck1999, Shukurov2003, Fletcher2006}. 
This effect is particularly important in extended sources
    such as emitting filaments in the cosmic web.
A simple Faraday screen 
    with a bright source behind a 
    Faraday-rotating medium 
    would be insufficient 
    to capture this effect properly. 
A covariant polarised radiative transfer calculation is therefore essential 
    to evaluate the line-of-sight depolarisation effect from all radiation processes at different redshifts
    \citep[see][]{Chan2019}. 

\vspace*{0.3cm}  

\begin{figure}
    \centering
    \includegraphics[width=0.4\textwidth]{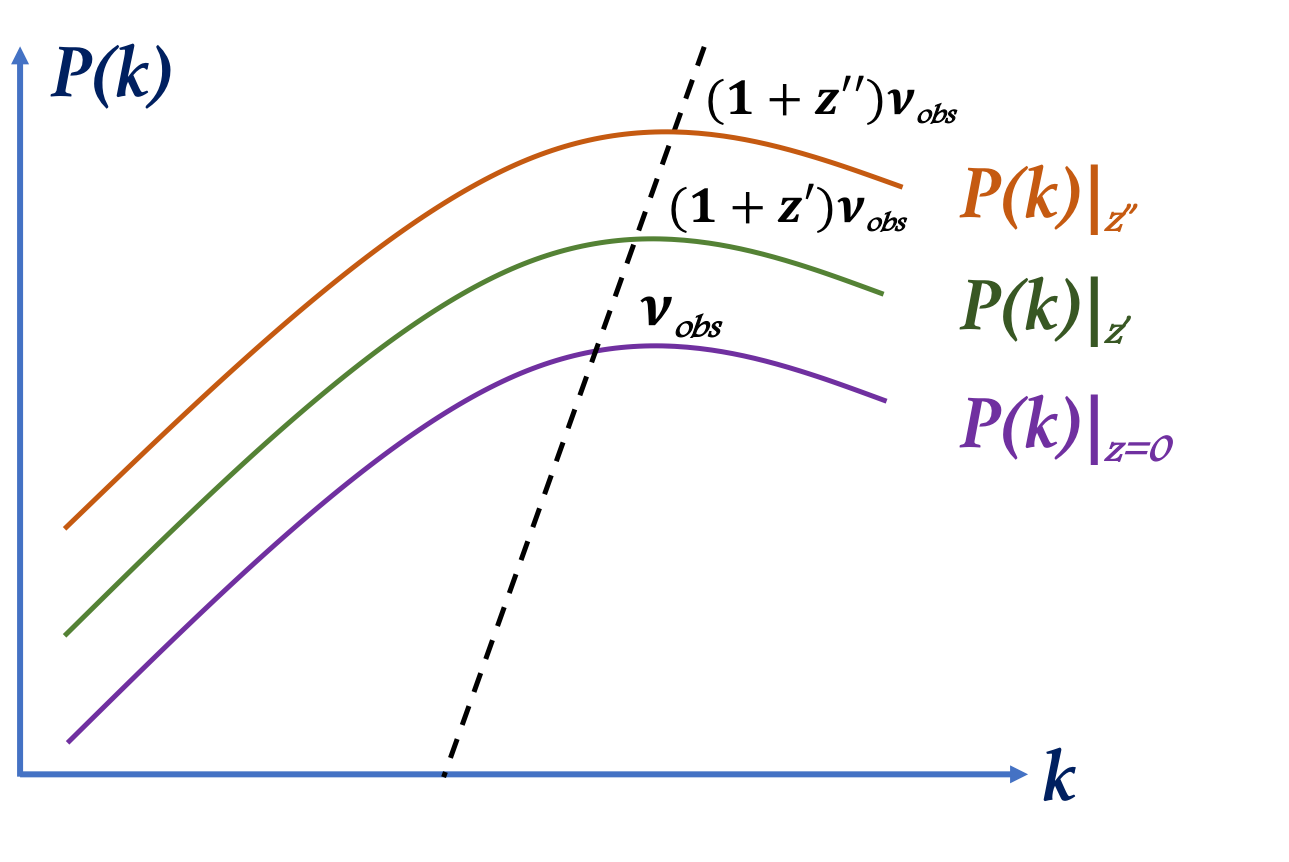}
    \caption{An illustration of how higher-redshift structures can contaminate the observed power spectrum at a fixed $\nu_{\rm obs}$ (dashed line),
        which differs from considering the theoretical power spectra $P(k)|_z$ from all sources at a single redshift.}
    \label{fig:powerspectrum}
\end{figure}

\noindent 
{\it Contamination in the power spectrum}:\\  
The power spectrum of the observed polarised intensity may be contaminated by emissions from the medium and embedded sources. 
Contributions from these sources would lead to 
    apparent higher power in fluctuations at small-scales. 
It is important to assess whether these signatures due to spatially separated sources can be distinguished from those imparted due to the true structures of magnetic fields. 
In addition to these, the interpretation of the power spectrum is 
    complicated by the contributions at various cosmological redshifts. 
Consider a radio observation of 
the sky
at a fixed frequency $\nu_{\rm obs}$.
The observed power spectrum $P_{k}$ is the result of 
contributions from sources at different redshifts.
Hence, 
at each $k$, the power spectrum is contaminated by different levels of emission from various sources at higher redshifts 
    (see Fig.~\ref{fig:powerspectrum}),
    which differ from the power spectrum of the Universe at every redshift, $P(k)|_z$.
Local $P(k)|_z$ does not contain any contribution from the higher redshifts,
    whereas observationally, different components at higher $k$ are picked up at $\nu_{\rm obs}$.


\section{Conclusions}
\label{sec-conclusions}

Faraday rotation measure fluctuation (RMF) analysis at radio wavelengths is considered as a diagnostic tool for cosmic magnetism.
Most of the current methods in RMF analyses rely on a random-walk model
in which the standard deviation of RMF provides a statistical measure of the field correlation length. 
Our objective is to assess the validity of the conventional random walk method as a cosmic magnetic field probe. 
We simulate various configurations of density and magnetic field fluctuations in astrophysical plasmas to calculate the dispersion of RM. 
We calculate and compare the line-of-sight longitudinal dispersion with the sky transverse dispersion.

Our results 
are as follows:
(i) Numerically,
    the divergence filtering also creates a residual dipole, as a result of IDL's preferential direction in its Fourier transform function.
This can be removed by taking a linear superposition of three orthogonal field realisations.
(ii) The conventional random walk model 
applies in some but not all astrophysical situations. 
More specifically, 
it is valid when the density fluctuations are uniformly-distributed or Gaussian-distributed.
The model breaks down for densities with fractal and log-normal structures.
(iii) Density fluctuations can obscure the effect of magnetic field fluctuations, and therefore affect the correlation length of magnetic fields determined by the conventional random walk model.
Our results show that
it is difficult to disentangle the signals from density and field fluctuations based on the value of the standard deviation of the RM, $\sigma_{_\mathcal{R}}$, itself.
More specifically,
our demonstration models show that
    different statistical indicators can potentially mislead,
    $\sigma_{_\mathcal{R}}<\varsigma_{_\mathcal{R}}$,
    by tens of percents or by factors of a few
    (if there is unrecognised cloudiness).

Even without degeneracy between the signals from density and field fluctuations, radiative processes such as absorption and emission 
can confuse and ambiguate 
the interpretation of the RM. 
Moreover, in addition to the thermal electrons, non-thermal electrons can also contribute to the Faraday rotation.
We conclude that the random walk approach is not universally valid and a more proper treatment based on (covariant) polarised radiative transfer 
in spatially detailed models is necessary 
    to develop solid theoretical models and predictions in preparation for the SKA.


\section*{Acknowledgments}  
We thank 
  Dr David Barnes, Dr Ziri Younsi, 
  Prof JinLin Han, Prof Bryan Gaensler, 
  Dr Jennifer West, Dr Cameron Van Eck, 
  Prof Kiyoshi Masui, Ryan Mckinven, 
  Dr Amit Seta, 
  Dr Shane O'Sullivan and
  Prof Kandaswamy Subramanian
  for helpful discussions. 
We also thank Dr Ellis Owen 
  for carefully reading the manuscript.
AYLO's visit to National Tsing Hua University 
  was hosted by Prof Albert Kong  
  and supported by the 
  Ministry of Science and Technology of the ROC (Taiwan)  
  grant 105-2119-M-007-028-MY3.  
She is also supported 
  by the Brunei Ministry of Education Scholarship. 
JYHC is supported by UCL through the GRS/ORS 
  and MAPS Deans Summer Research Studentship, 
  and by Lady Margaret Hall, University of Oxford through a research scholarship. 
KW thanks the hospitality of Perimeter Institute where part of this work was carried out. Research at Perimeter Institute is supported in part by the Government of Canada through the Department of Innovation, Science and Economic Development Canada and by the Province of Ontario through the Ministry of Economic Development, Job Creation and Trade. 
LVDG is partially funded under STFC consolidated grant No. ST/S000240/1. 
A part of this work was undertaken on ARC3,
part of the High Performance Computing facilities
at the University of Leeds, UK.
This research had made use of NASA's Astrophysics Data System. 


\bibliography{reference_RM} 
\bibliographystyle{mnras}

\appendix

\section{Divergence-free filter for the magnetic field} 
\label{app-div_free}

Consider a vector ${\boldsymbol k} \, (=k_{\rm i})$,
  defining a reference axis in a vector space.
Any other arbitrary vector ${\boldsymbol X}$
  can be decomposed into two components, 
  one parallel to and another one perpendicular to ${\boldsymbol k}$: 
 ${\boldsymbol X} = {\boldsymbol X}^{(k)}_\parallel + {\boldsymbol X}^{(k)}_\perp$, 
  with $\big|\;\!{\boldsymbol k} \cdot  {\boldsymbol X}^{(k)}_\parallel  \big| \geq 0$ 
  and ${\boldsymbol k} \cdot  {\boldsymbol X}^{(k)}_\perp = 0$. 
Now introduce a projection operator ${\cal P} ({\boldsymbol k})$, 
  such that,  ${\boldsymbol X}' = {\cal P} ({\boldsymbol k})\;\! {\boldsymbol X}  
  =  {\boldsymbol X}^{(k)}_\perp$. 
This projection operator
  eliminates the longitudinal component of ${\boldsymbol X}$, 
  ensuring that ${\boldsymbol k} \cdot {\boldsymbol X}' = 0$ 
  for any given ${\boldsymbol X}\,$. 
A non-trivial example
 of ${\cal P} ({\boldsymbol k})$
 is
 $({\boldsymbol I} -{\boldsymbol {\hat k}}{\boldsymbol {\hat k}})$,  
 where ${\boldsymbol I}$ is an identity operator and 
 ${\boldsymbol{\hat  k}} = {\boldsymbol k}/|{\boldsymbol k}|$, such that,    
\begin{equation} 
 {\boldsymbol k} \cdot \left[ \big({\boldsymbol I} 
 - {\boldsymbol{\hat k}}{\boldsymbol{\hat k}} \big)
    \cdot {\boldsymbol X} \right]  \ = \ 
     k_{\rm i} \left(\delta_{\rm ij} 
      - \frac{k_{\rm i}k_{\rm j}}{k^2} \right) X_{\rm j} 
      \  = \  0 \ . 
\end{equation} 
   
Magnetic fields in vacuum are solenoidal, i.e.\ divergence-free, 
  satisfying $\nabla \cdot {\boldsymbol B} 
   = {\partial\;\!}_{\rm i} B_{\rm i} = 0$. 
In Fourier space, the divergence-free relation is expressed as   
   ${\boldsymbol k} \cdot {\boldsymbol B}({\boldsymbol k}) = k_{\rm i} B_{\rm i} = 0$, 
   which requires that the field component parallel to 
   ${\boldsymbol k}$ must vanish. 
Thus, we may apply the filter 
$({\boldsymbol I} - {\boldsymbol{\hat k}}{\boldsymbol{\hat k}})$ 
  in Fourier space 
  to prepare a divergence-free magnetic field
  (with designated structural properties)
  from a generic initial simulated random vector field
  (with otherwise the same structural properties). 
The procedures are as follows:\\
(i) 
Construct a random field 
${\boldsymbol {\tilde B}}({\boldsymbol k})\, 
  (= {\tilde B}_{\rm i}(k_{\rm m}))$ 
  according to the specified structural properties in Fourier space.\\ 
(ii) 
Apply the divergence-free filter, 
i.e.\ carry out the projection operation: 
   $B_{\rm i}(k_{\rm m}) = (\delta_{ij} - k_{\rm i}k_{\rm j}/k^2) {\tilde B}_{\rm j}(k_{\rm m})$.\\ 
(iii) 
Use an inverse-Fourier transform on $B_{\rm i}(k_{\rm m})$ 
      to obtain $B_{\rm i}(x_{\rm m})$ in configuration space.\\
As the filtering process removes the longitudinal part of the magnetic field in Fourier space, 
  it reduces the total magnetic energy stored in the system. 
The Parseval's (energy) Theorem,   
\begin{equation}  
   \int_{V_x} d^3{\boldsymbol x} \ 
   \big| {\boldsymbol B} ({\boldsymbol x}) \big|^2  
   =
  \int_{V_k} d^3{\boldsymbol k} \ 
  \big| {\boldsymbol B} ({\boldsymbol k})   \big|^2     
  \  ,      
\end{equation} 
  requires that the total magnetic energy 
  is reduced by the same amount 
  in configuration space as in Fourier space. 
With the divergence-free magnetic field given by 
${\boldsymbol B}({\boldsymbol k}) = 
   ({\boldsymbol I} -  
   {\boldsymbol{\hat  k}}{\boldsymbol{\hat k}})\, 
  {\boldsymbol {\tilde B}}({\boldsymbol k})$,  
   the energy density of the magnetic field is   
\begin{align}
    \frac{1}{8\uppi}\,   \left|{\boldsymbol B}\right|^2 & =  \frac{1}{8\uppi} \left[   \left( \delta_{\rm ij} - \frac{k_{\rm i}k_{\rm j}}{k^2}  \right) 
      \left( \delta_{\rm im} - \frac{k_{\rm i}k_{\rm m}}{k^2}  \right) 
       {\tilde B}_{\rm j} {\tilde B}_{\rm m}    \right]  \nonumber \\ 
      & =  \frac{1}{8\uppi} \left[  {\tilde B}_{\rm i} {\tilde B}_{\rm i} - \frac{1}{k^2} \left(  k_{\rm i} {\tilde B}_{\rm i}\right)^2 \right]  \nonumber \\ 
      & = \frac{1}{8\uppi} \,  \left|{\tilde {\boldsymbol B}}\right|^2 \, (1-\mu^2) \ , 
\end{align} 
    where $\mu = {\boldsymbol{\hat k}}\cdot 
     {\boldsymbol{\tilde  B}}/|{\boldsymbol{\tilde B}}|$. 
For a randomly-oriented magnetic field in Fourier space, 
\begin{equation} 
  \big\langle 1- \mu^2 \big\rangle 
  = \frac{1}{2} \int_{-1}^1 d\mu \  
   \left(1-\mu^2\right)  \ \ = \ \ \frac{2}{3}  \ .   
\end{equation} 
Hence, one-third of the magnetic energy density is filtered out. 
This is the expected amount   
  when there is equipartition between the energies in  
  the longitudinal component and the two orthogonal perpendicular components 
  (the solenoidal components) 
  of the initial ``magnetic'' field ${\boldsymbol {\tilde B}}$.  
To recover the energy loss in the divergence-free filtering process, 
  we may renormalise the resulting divergence-free magnetic field,  
  either in configuration space or in Fourier space,  
  by a multiplicative factor $\sqrt{3/2}\,$.

\section{Preparation of the model 2-phase fractal clouds} 
\label{app-cloudy_blocks}

Starting from even the most minute inhomogeneities,
   astrophysical plasmas are susceptible to form substructures
   through a variety of thermal, magnetic, and buoyancy instabilities
   \citep[e.g.][]{Field1965,Shu1972,Balbus1989,Quataert2008,McCourt2012,Sharma2012,Wareing2016}.
An optically thin plasma of nearly solar composition
   has a temperature-dependent radiative cooling function
   that incurs thermal instability over an interval
   $10^4\,\mathrm{K}\la T\la10^7\,\mathrm{K}$.
An initially homogeneous medium can spontaneously self-segregate
   into a quasi-equilibrium of two coexisting phases:
   the original hot diffuse medium;
   and a minor component of cooler dense clouds.
Externally imposed isobaric conditions imply a density ratio $\ga10^3$
   between the phases,
   in the absence of any further gravitational collapse.
Thermally condensed clouds are endemic in otherwise hot extragalactic media,
   and can stretch into filaments in upflows and downflows
   associated with active galaxies
   \citep[e.g.][]{Ford1979,Saxton2001,Conselice2001,McDonald2010,
        Voit2017,Combes2018,Olivares2019}.

As a test of RMF due to strong density inhomogeneities,
   we build two-phase toy models
   capable of approximating the knotty medium of a galaxy cluster core,
   or the ISM of an elliptical galaxy that acquired clouds
   (either via thermal instability or a wet-dry merger).
Initially
   we generate a Gaussian distribution of pseudo-random complex numbers,
   and apply an amplitude filter to obtain
   a Kolmogorov-like power spectrum.
This cube is transformed according to the Elmegreen recipe
   for imitating lognormal density fluctuations in a turbulent medium,
   which will represent the diffuse phase.
We prescribe a volume filling factor of clouds ($0<f\ll1$)
   and select the densest ranked voxels, 
   down to a suitable threshold.
Their densities are multiplied by a uniform constant,
   set to ensure a mean density ratio of $10^3$
   between the cloud and non-cloud phases.
Assuming that the clouds are condensing from the hot medium,
   we normalise the mean of the entire cloudy block to $10^{-3}\,\mathrm{cm}^{-3}$,
   matching the standard for our single-phase density models.

We create and test models ranging from
   a negligible smattering of clouds ($f\approx10^{-6}$)
   to a heavily obscured overcast case ($f\approx10^{-2}$)
   where a majority ($\ga 0.7$) of RM map pixels or rays
   traverse at least one dense cloud.
Table~\ref{tab:app-cloudy}
   presents basic global properties of these models.
In area terms, the cloud coverage factors decrease with $f$,
   and vary with orientation due to the clouds' random fractal shapes.
Clouds account for only a tiny fraction of the total mass
   in Cd4--Cd6,
   or just under half the mass in Cd3.
The overcast case Cd2 is dominated by the mass of the dense cold phase,
   making it unrealistic for the filament-infused core of a galaxy cluster
   (where the cold fraction is at most a few tens of percents),
   but perhaps more like the primordial medium
   of a hypothetical wet protogalaxy.
The mean densities of the cubes vary by factors of a few
   before their normalisations into fiducial ICM units.

\begin{table}
    \centering
    \begin{tabular}{clllllllll}
\hline
\\
\multicolumn{1}{c}{model}
&$\log f$
&$m_\mathrm{c}/M$
&$a_x/A$
&$a_y/A$
&$a_z/A$
\\
\\
\hline
\\
%
Cd2   &-2.01    &0.912           &0.713       &0.752       &0.749
\\
Cd3   &-3.01    &0.495           &0.150       &0.155       &0.155
\\
Cd4   &-4.02    &0.0877          &0.0188      &0.0190      &0.0189
\\
Cd5   &-5.12    &0.00746         &0.00171     &0.00172     &0.00169
\\
Cd6   &-6.32    &0.000477        &0.000122    &0.000122    &0.000122
\\
\\
\hline
    \end{tabular}
    \caption{%
    Summary of cloudy model properties:
    volume filling factors $f$;
    cloud mass fraction;
    area covering factors for
    the three orthogonal views.
    Before radiative transfer calculations,
    all models are normalised to the same total mass
    or mean density.}
    \label{tab:app-cloudy}
\end{table}

\bsp	
\label{lastpage}
\end{document}